\documentclass{aa}

\usepackage{graphicx,url,twoopt,natbib}
\usepackage[varg]{txfonts}
\usepackage[pdftitle={Conditions for water ice lines and Mars-mass exomoons around accreting super-Jovian planets at 1 - 20\,AU from Sun-like stars}, colorlinks = true, breaklinks = true, citecolor = blue, linkcolor = blue, urlcolor = blue, pdfauthor = {Ren\'{e} Heller}]{hyperref}
\usepackage{breakurl}
\bibpunct{(}{)}{;}{a}{}{,}    

\begin{document} 

\title{Conditions for water ice lines and Mars-mass exomoons around accreting super-Jovian planets at 1 - 20\,AU from Sun-like stars}

\titlerunning{Water ice lines and Mars-mass exomoons around accreting super-Jovian planets at 1 - 20\,AU}

\author{R. Heller\inst{1}\fnmsep\thanks{Postdoctoral Fellow of the Canadian Astrobiology Training Program}
            \and
            R. Pudritz\inst{1}
           }

\institute{Origins Institute, McMaster University, 1280 Main Street West, Hamilton, ON L8S 4M1, Canada \\
              \email{rheller@physics.mcmaster.ca} , \email{pudritz@physics.mcmaster.ca}
              }

\date{Received December 9, 2014; accepted April 4, 2015}
 
\abstract
{The first detection of a moon around an extrasolar planet (an ``exomoon'') might be feasible with NASA's \textit{Kepler} or ESA's upcoming \textit{PLATO} space telescopes or with the future ground-based \textit{European Extremely Large Telescope}. To guide observers and to use observational resources most efficiently, we need to know where the largest, most easily detected moons can form.}
{We explore the possibility of large exomoons by following the movement of water (H$_2$O) ice lines in the accretion disks around young super-Jovian planets. We want to know how the different heating sources in those disks affect the location of the H$_2$O ice lines as a function of stellar and planetary distance.}
{We simulate 2D rotationally symmetric accretion disks in hydrostatic equilibrium around super-Jovian exoplanets. The energy terms in our semi-analytical framework -- (1) viscous heating, (2) planetary illumination, (3) accretional heating of the disk, and (4) stellar illumination -- are fed by precomputed planet evolution models. We consider accreting planets with final masses between 1 and 12 Jupiter masses at distances between 1 and 20\,AU to a solar type star.}
{Accretion disks around Jupiter-mass planets closer than about 4.5\,AU to Sun-like stars do not feature H$_2$O ice lines, whereas the most massive super-Jovians can form icy satellites as close as 3\,AU to Sun-like stars. We derive an empirical formula for the total moon mass as a function of planetary mass and stellar distance and predict that super-Jovian planets forming beyond about 5\,AU can host Mars-mass moons. Planetary illumination is the major heat source in the final stages of accretion around Jupiter-mass planets, whereas disks around the most massive super-Jovians are similarly heated by planetary illumination and viscous heating. This indicates a transition towards circumstellar accretion disks, where viscous heating dominates in the stellar vicinity. We also study a broad range of circumplanetary disk parameters for planets at 5.2\,AU and find that the H$_2$O ice lines are universally between about 15 and 30 Jupiter radii in the final stages of accretion when the last generation of moons is supposed to form.}
{If the abundant population of super-Jovian planets around 1\,AU formed in situ, then these planets should lack the previously predicted population of giant icy moons, because those planets' disks did not host H$_2$O ice in the final stages of accretion. But in the more likely case that these planets migrated to their current locations from beyond about 3 to 4.5\,AU they might be orbited by large, water-rich moons. In this case, Mars-mass ocean moons might be common in the stellar habitable zones. Future exomoon detections and non-detections can provide powerful constraints on the formation and migration history of giant exoplanets.}

\keywords{Accretion, accretion disks -- Planets and satellites: formation -- Planets and satellites: gaseous planets -- Planets and satellites: physical evolution -- Astrobiology}

\maketitle

\section{Introduction}

Now that the detection of sub-Earth-sized objects has become possible with space-based photometry \citep{2012ApJ...747..144M,2013Natur.494..452B} and almost 2000 extrasolar planets have been confirmed \citep{2013ApJS..204...24B,2014ApJ...784...45R}, technological and theoretical advances seem mature enough to find moons orbiting exoplanets. Natural satellites similar in size to Mars ($0.53$ Earth radii, $R_\oplus$) or Ganymede ($0.41\,R_\oplus$) could be detectable in the available \textit{Kepler} data \citep{2012ApJ...750..115K,2014ApJ...787...14H}, with the upcoming \textit{PLATO} space mission \citep{2012MNRAS.419..164S,2014ExA...tmp...41R} or with the \textit{European Extremely Large Telescope} (\textit{E-ELT}) \citep{2014ApJ...796L...1H}. Exomoons at about 1\,AU or closer to their star might be detected during stellar transits \citep{1999A&AS..134..553S,2006A&A...450..395S,2009MNRAS.400..398K}. Young, self-luminous planets beyond 10\,AU might reveal their satellites in the infrared through planetary transits of their moons, if the unresolved planet-moon binary can be directly imaged \citep{2013ApJ...769...98P,2014ApJ...796L...1H}. Moon formation theories can guide observers and data analysts in their searches to streamline efforts. In turn, the first non-detections of exomoons \citep{2001ApJ...552..699B,2007A&A...476.1347P,2013ApJ...770..101K,2013ApJ...777..134K,2014ApJ...784...28K} and possible future findings provide the first extrasolar observational constraints on satellite formation.

In a recent study, we developed a circumplanetary disk accretion model and simulated the radial motion of the water (H$_2$O) ice line around super-Jovian\footnote{Throughout the paper, our reference to ``super-Jovian'' planets includes planets with masses between that of Jupiter ($M_{\rm Jup}$) and $12\,M_{\rm Jup}$, where the latter demarcates the transition into the brown dwarf regime.} exoplanets at 5.2\,AU around Sun-like stars \citep{2014arXiv1410.5802H}. The H$_2$O ice line is critical for the formation of large, possibly detectable moons, because here the surface density of solids increases sharply by a factor of 3 to 4 \citep{1981PThPS..70...35H}. The composition and the masses of the Galilean moons are usually considered as records of the location of the H$_2$O ice line and, more generally, of the temperature distribution in Jupiter's accretion disk \citep{1974Icar...21..248P,1982Icar...52...14L}. The inner moons Io and Europa turned out mostly rocky and comparatively light, whereas Ganymede and Callisto formed extremely rich in water ices (about 50\,\%) but became substantially more massive \citep{1999Sci...296...77S}.

Our main findings were that (1) super-Jovian planets at 5.2\,AU around Sun-like stars have their H$_2$O ice lines between about 15 and 30 Jupiter radii ($R_{\rm Jup}$) during the late stages of accretion, when the final generation of moons form. This range is almost independent of the planetary mass ($M_{\rm p}$). (2) With the most massive planets having the most widely extended disks, these disks host the largest reservoirs of H$_2$O ices. In particular, the total instantaneous mass of solids ($M_{\rm sld}$) in these disks scales roughly proportional with $M_{\rm p}$. (3) The current orbital position of Ganymede is very close the mean radial location of the circumjovian H$_2$O ice line in our simulations, suggesting a novel picture in which Ganymede formed at a circumplanetary ice line trap. (4) Heat transitions, however, transverse the accretion disks on a very short timescale ($\approx10^4$\,yr) once the planet opens up a gap in the circumstellar accretion disk and thereby drastically reduces the supply of material. This timescale is short compared to the satellite migration time scale ($10^5$ - $10^7$\,yr) \citep{2002ApJ...565.1257T,2002AJ....124.3404C,2006Natur.441..834C,2010ApJ...714.1052S,2012ApJ...753...60O} and the time that is left for the final accretion until shutdown ($10^5$\,yr). Hence, heat transitions around super-Jovian planets cannot act as moons traps, which indicates a different behavior of planet and moon formation \citep{2004ApJ...606..520M,2011MNRAS.417.1236H,2012ApJ...755...74K}.

\begin{figure}[t]
   \centering
   \scalebox{0.605}{\includegraphics{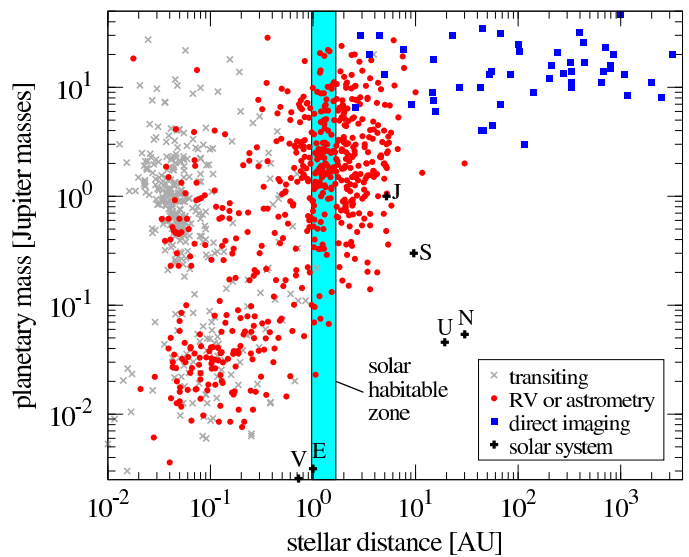}}
   \caption{Stellar distances and planetary masses of extrasolar planets listed on \url{www.exoplanet.eu} as of 5 April 2015. Symbols indicate the discovery method of each planet, six solar system planets are shown for comparison. Note the cluster of red dots around 1\,AU along the abscissa and between 1 and $10\,M_{\rm Jup}$ along the ordinate. These super-Jovian planets might be hosts of Mars-mass moons. The shaded region denotes the solar habitable zone defined by the runaway and maximum greenhouse as per \citet{2013ApJ...765..131K}.}
   \label{fig:exoplanets}
 \end{figure}

We here extend the parameter space of our previous paper and consider planetary accretion between 1 and 20\,AU from the star. This range is motivated by two facts. First, super-Jovians are the most abundant type of confirmed planets, that is, objects with accurate mass estimates at these distances (see Fig.~\ref{fig:exoplanets})\footnote{Data from the \textit{Kepler} space telescope suggests, however, that terrestrial planets are more abundant at about 1\,AU around Sun-like stars than gas giant planets \citep{2013ApJ...778...53D,2015ApJ...798..112M}.}. And second, this range contains the stellar habitable zone around Sun-like stars, which is of particular interest given the fact that giant, water-rich moons might be frequent habitable environments \citep{2014AsBio..14...50H,2014AsBio..14..798H}. Our main goal is to locate the H$_2$O ice lines in the circumplanetary accretion disks at the time of moon formation shutdown. Their radial separations from the planet will correspond to the orbital radii at or beyond which we can suspect the most massive, water-rich moons to reside. In particular, stellar illumination at only a few AU from the star will prevent the accretion disks from having H$_2$O ice lines and therefore from hosting large moons. We determine these critical stellar distances below.

\renewcommand*{\arraystretch}{1.35}
\setlength{\tabcolsep}{5pt}
\begin{table*}[]
\begin{center}
\caption{Parameterization of the circumplanetary disk as described in \citet{2014arXiv1410.5802H}.}
\begin{tabular}{ccc}
\hline\hline
Symbol & Meaning & Fiducial Value\\
\hline\hline
& Constant or parameterized planetary disk parameters & \\
\hline
$r$ & distance to the planet &  \\
$R_\mathrm{H}$ & planetary Hill radius & \\
$r_\mathrm{cf}$ & centrifugal radius & \\
$r_\mathrm{c}$ & transition from optically thick to optically thin (viscously spread) outer part & $27.22 / 22 \times r_\mathrm{cf}$ \tablefoottext{a} \\
$r_\mathrm{d}$ & outer disk radius & $R_\mathrm{H}/5$\tablefoottext{b,c} \\
$\Lambda(r)$ & radial scaling of $\Sigma(r)$ & \\
$T_\mathrm{m}(r)$ & midplane temperature & \\
$T_\mathrm{s}(r)$ & surface temperature & \\
$\Sigma(r)$ & gas surface density &  \\
$h(r)$ & effective half-thickness & \\
$\rho_0(r)$ & gas density in the midplane & \\
$\rho_\mathrm{s}(r)$ & gas density at the radiative surface level & \\
$z_\mathrm{s}(r)$ & radiative surface level, or photospheric height & \\
$q_\mathrm{s}(r)$ & vertical mass coordinate at the radiative surface & \\
$c_\mathrm{s}(T_\mathrm{m}(r))$ & speed of sound &  \\
$\alpha$ & Viscosity parameter & 0.001\tablefoottext{d} \\
$\nu(r,T_\mathrm{m})$ & gas viscosity & \\
$\Gamma$ & adiabat exponent, or ratio of the heat capacities & 1.45 \\
$\chi$ & dust enrichment relative to the protostellar, cosmic value & 10\tablefoottext{c} \\
$X_\mathrm{d}$ & dust-to-mass fraction & $0.006$\tablefoottext{e,f}\\
$\mu$ & mean molecular weight of the H/He gas & $2.34\,\mathrm{kg/mol}$ \tablefootmark{\textit{(g)}}\\
\hline
& Variable planetary disk parameters & \\
\hline
$\kappa_\mathrm{P}$ & Planck mean opacity & $10^{-3} \ - \ 10^{-1}\,\mathrm{m}^2\,\mathrm{kg}^{-1}$\tablefoottext{h}\\
$k_\mathrm{s}$ & fraction of the solar radiation flux contributing to disk heating at $z \leq z_\mathrm{s}$ & $0.1 - 0.5$\tablefoottext{c} \\
$\dot{M}_\mathrm{shut}$ & shutdown accretion rate for moon formation & $100, 10, 1 \, M_{\rm Gan}/\mathrm{Myr}$ \\
\hline
\end{tabular}\label{tab:parameters}
\end{center}
\tablefoot{See \citet{2014arXiv1410.5802H} and additional references for details about the respective parameter values or ranges: \tablefoottext{a}{\citet{2008ApJ...685.1220M}} \tablefoottext{b}{\citet{2010ApJ...714.1052S}} \tablefoottext{c}{\citet{2014SoSyR..48...62M}} \tablefoottext{d}{\citet{2014MNRAS.440...89K}} \tablefoottext{e}{\citet{1982Icar...52...14L}} \tablefoottext{f}{\citet{2013ApJ...778...78H}} \tablefoottext{g}{\citet{2013ApJ...778...77D}} \tablefoottext{h}{\citet{1997ApJ...486..372B}}.}
\end{table*}

 \begin{figure*}[t]
   \centering
   \scalebox{0.61}{\includegraphics{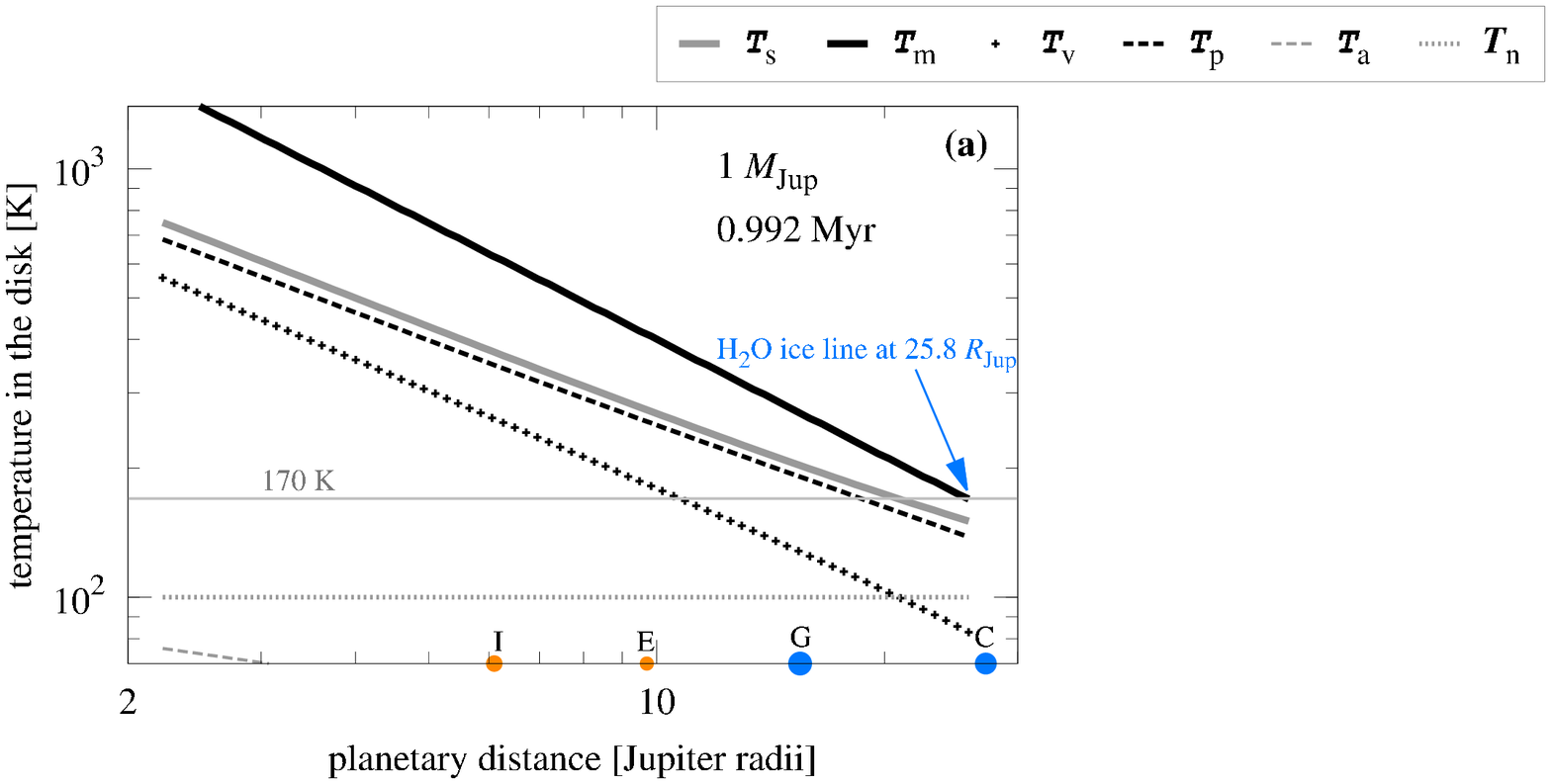}}
    \hspace{-3.7cm}
   \scalebox{0.61}{\includegraphics{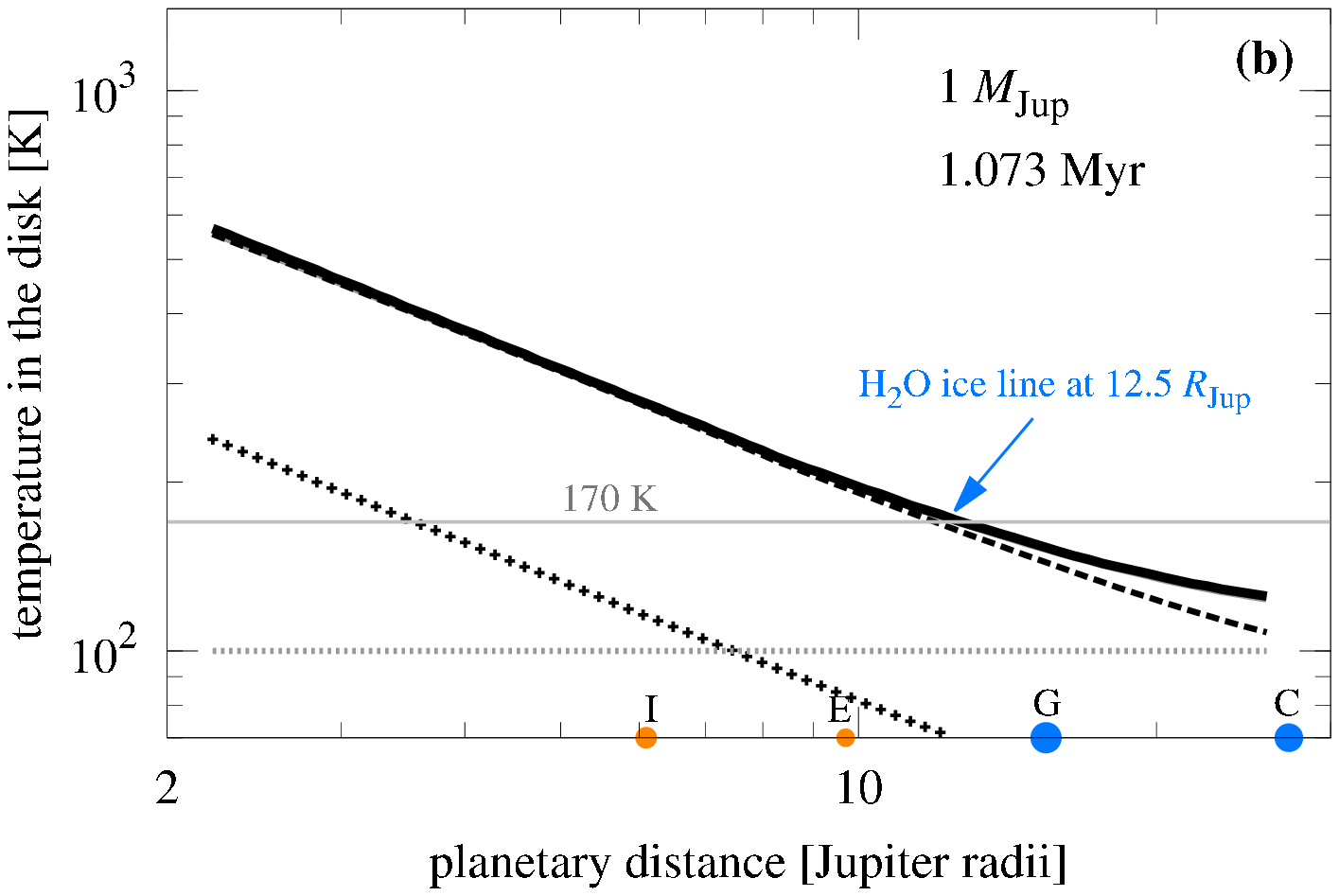}}
   \caption{Temperature structure in the disk around a Jupiter-mass planet 5.2\,AU from a Sun-like star. Solid black lines indicate disk midplane temperatures, solid gray lines disk surface temperatures. The other lines indicate a hypothetical disk surface temperature assuming only one heat source (see legend).  \textbf{(a)}: At 0.992\,Myr in this particular simulation, the disk has sufficiently cooled to allow the appearance of an H$_2$O ice line in the disk midplane at the outer disk edge at $25.8\,R_{\rm Jup}$. The planetary accretion rate is about $4~\times10^2\,M_{\rm Gan}\,{\rm Myr}^{-1}$. \textbf{(b)}: At 1.073\,Myr, when the planetary accretion rate has dropped to $10\,M_{\rm Gan}\,{\rm Myr}^{-1}$, the H$_2$O ice line has settles between the orbits of Europa and Ganymede (see colored symbols). In these final stages of accretion, disk temperatures are governed by planetary illumination (see dashed line for $T_{\rm p}$).}
   \label{fig:T_shots_1}
 \end{figure*}

 \begin{figure*}[t]
   \centering
   \scalebox{0.61}{\includegraphics{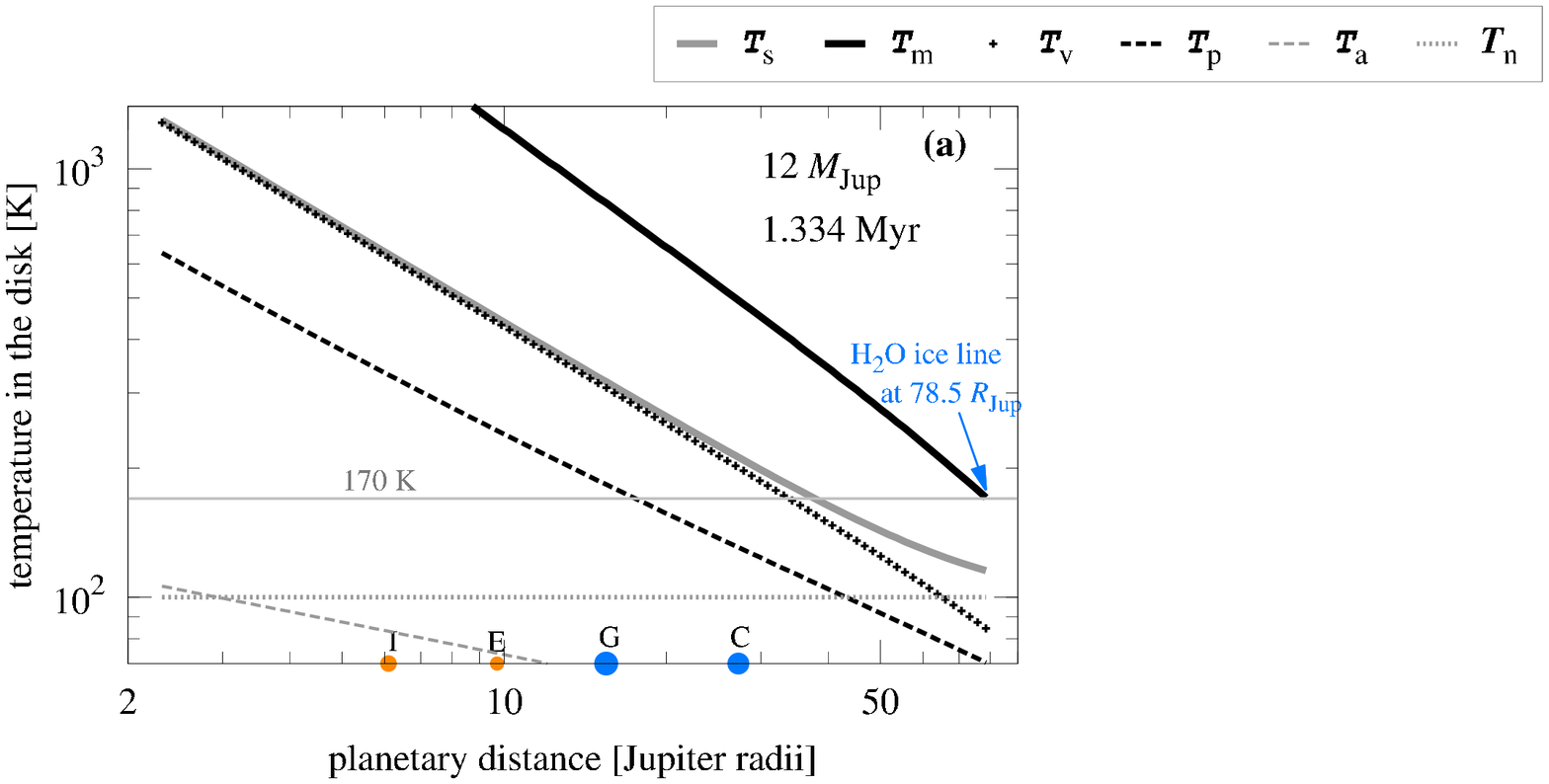}}
    \hspace{-3.7cm}
   \scalebox{0.61}{\includegraphics{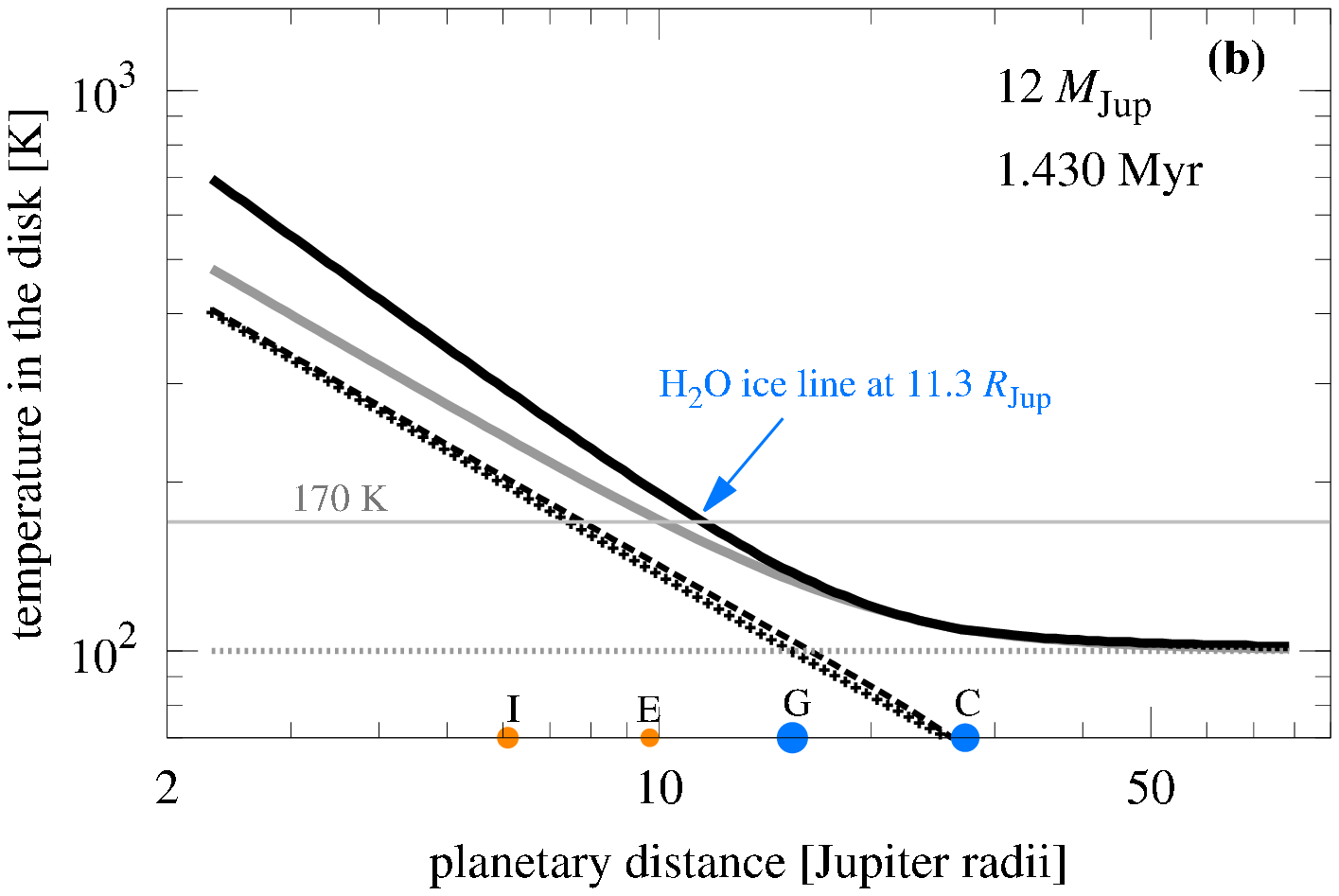}}
   \caption{Same as Fig.~\ref{fig:T_shots_1}, but now for a $12\,M_{\rm Jup}$ planet. Note that the accretion disk is substantially larger than in the $1\,M_{\rm Jup}$ case. \textbf{(a)}: The H$_2$O ice line appears much later than in the Jovian scenario, here at 1.334\,Myr. The planetary accretion rate is about $10^3\,M_{\rm Gan}\,{\rm Myr}^{-1}$ and viscous heating is dominant in this phase. \textbf{(b)}: Once the planetary accretion rate has dropped to $10\,M_{\rm Gan}\,{\rm Myr}^{-1}$ at $1.43$\,Myr, the H$_2$O ice line is at a similar location as in the Jupiter-like scenario, that is, between the orbits of Europa and Ganymede.}
   \label{fig:T_shots_12}
 \end{figure*}

\section{Methods}

We use the framework developed in \citet{2014arXiv1410.5802H}, which models a 2D axisymmetric accretion disk in hydrostatic equilibrium around the planet. It considers four heating terms, or energy fluxes, of the disk, namely (1) viscous heating ($F_{\rm vis}$), (2) planetary illumination ($F_{\rm p}$), (3) accretional heating of the disk ($F_{\rm acc}$), and (4) stellar illumination, all of which determine the midplane and surface temperature of the disk as a function of planetary distance ($r$). This model \citep[based on earlier work by][]{1995SoSyR..29...85M,2014SoSyR..48...62M} considers $F_{\rm vis}$ as a distributed energy source within the disk, while $F_{\rm p}$, $F_{\rm acc}$, and stellar illumination are considered external heat sources. The model also includes an analytical treatment for the vertical radiative energy transfer, which depends on the Planck mean opacity ($\kappa_{\rm P}$). In real disks, $\kappa_{\rm P}$ will depend on the disk temperature and the composition of the solids. In particular, it will be a function of the radial distance to the planet and it will evolve in time as small particles stick together and coagulate. To reduce computational demands, we here assume a constant $\kappa_{\rm P}$ throughout the disk, but we will test values over two orders of magnitude to explore the effects of changing opacities. The vertical gas density in the disk is approximated with an isothermal profile, which is appropriate because we are mostly interested in the very final stages of accretion when the disk midplane and the disk surface have similar temperatures. Furthermore, deviations between an isothermal and an adiabatic vertical treatment are significant only in the very dense and hot parts of the disk inside about $10\,R_{\rm Jup}$. As the H$_2$O ice line will always be beyond these distances, inaccuracies in our results arising from an isothermal vertical model are negligible.

The heating terms (1)-(3) are derived based on precomputed planet evolution models \citep[provided by courtesy of C.~Mordasini,][]{2013A&A...558A.113M}, which give us the planetary mass, planetary mass accretion rate ($\dot{M}$), and planetary luminosity ($L_{\rm p}$) as a function of time ($t$) \citep[see Fig.~1 in][]{2014arXiv1410.5802H}. $L_{\rm p}(t)$ is a key input parameter to our model as it determines $F_{\rm p}(t)$, and it is sensitive to the planet's core mass. Yet, in the final stages of planetary accretion, $L_{\rm p}$ differs by less than a factor of two for planetary cores between 22 to $130$ Earth masses. With $F_{\rm p}~\propto~L_{\rm p}$ and temperature scaling roughly with $(F_{\rm p})^{1/4}$, uncertainties in the midplane or surface temperatures of the disk are lower than $20\,\%$, and uncertainties in the radial distance of the H$_2$O ice line are as large as a few Jovian radii at most. For all our simulations the precomputed planetary models assume a final core mass of 33 Earth masses, which is about a factor of three larger than the mass of Jupiter's core \citep{1997Icar..130..534G}. With higher final core masses meaning higher values for $L_{\rm p}$ at any given accretion rate, we actually derive upper, or outer, limits for the special case of the H$_2$O ice line around a Jupiter-like planet 5.2\,AU around a Sun-like star.

We interpolate the \citet{2013A&A...558A.113M} tracks on a linear scale with a time step of 1\,000\,yr and evaluate the four heating terms as a function of $r$, which extends from Jupiter's co-rotation radius \citep[$2.25\,R_{\rm Jup}$ in our simulations,][]{2010ApJ...714.1052S} out to the disk's centrifugal radius ($r_{\rm cf}$). At that distance, the centrifugal force acting on a gas parcel equals the gravitational pull of the planet. We compute $r_{\rm cf}$ using the analytical expression of \citet{2008ApJ...685.1220M}, which they derived by fitting a power law expression to their 3D hydrodynamical simulations of circumplanetary accretion disks. In this model, $r_{\rm cf}~\propto~M_{\rm Jup}^{1/3}$ for super-Jovian planets at a given stellar distance. Viscous heating is governed by the $\alpha$ viscosity parameter \citep{1973A&A....24..337S}, which we fix to a value of $10^{-3}$ in our simulations \citep[for a discussion see Sect.~4 in][]{2014arXiv1410.5802H}. Note that even variations of $\alpha$ by an order of magnitude would only change our results for the circumplanetary H$_2$O ice line location during the final stages of planetary accretion by a few planetary radii at most, because then the disk is mostly heated by planetary illumination.

The sound velocity ($c_{\rm s}$) in the disk midplane is evaluated as $c_{\rm s}~=~1.9\,{\rm km}\,{\rm s}^{-1}\sqrt{T_{\rm m}(r)/1\,000\,{\rm K}}$ \citep{2014MNRAS.440...89K}, which is an adequate approximation since ionization can be neglected in the late stages of moon formation when disk temperatures are usually below 1\,000\,K. At each time step, we assume a steady-state gas surface density ($\Sigma_{\rm g}$) that is derived analytically by solving the continuity equation of the mass inflow onto a centrifugally supported, viscous disk with a uniform flux per area \citep{2002AJ....124.3404C}. The equations of energy transport \citep{2014SoSyR..48...62M} then allow us to derive the temperature profile both at the disk surface, where the optical depth $\tau~=~2/3$, and in the disk midplane. This model invokes absorption of planetary illumination in the disk photosphere at a height $z_{\rm s}(r)$ above the midplane, modeled by an absorption coefficient ($k_{\rm s}$), as well as the transport of energy through an optically thick disk to the surface, modeled by $\kappa_{\rm P}$. For the dust-to-gas ratio ($X_\mathrm{d}$) we take a fiducial value of 0.006 \citep{1982Icar...52...14L,2013ApJ...778...78H} in the inner dry regions of the disk, and we assume that it jumps by a factor of three at the H$_2$O ice line. The resulting mathematical framework contains several implicit functions, which we solve in an iterative, numerical scheme. Table~\ref{tab:parameters} gives a complete overview of the parameters involved in the model. Our model neglects the effects of planetary migration, which is an adequate approximation because the planets don't migrate substantially within the $\approx10^{5}$\,yr required for moon formation. We also do not actually simulate the accretion and buildup of moons \citep{2012ApJ...753...60O}.

As an extension of our previous study, where all planets were considered at 5.2\,AU around a Sun-like star, we here place hypothetical super-Jovian planets at stellar distances between 1 and 20\,AU to a solar type host. Although the precomputed planet formation tracks were calculated at 5.2\,AU around a Sun-like star, we may still consider different stellar distances because the accretion rates through any annulus in the circumstellar disk are roughly constant at any time (and thereby similar to the instantaneous stellar accretion rate). The surface densities of gas and solids are naturally lower at larger stellar separations. Hence, the accretion rates provided by the tracks will actually overestimate $\dot{M}$ and $L_{\rm p}$ in wider orbits at any given time. However, this will not affect our procedure as we are not primarily interested in the evolution as a function of absolute times but rather as a function of accretion rates. In particular, we introduced the shutdown accretion ($\dot{M}_{\rm shut}$) in \citet{2014arXiv1410.5802H}, which we use as a measure to consider as the final stages of moon formation around accreting giant planets. $\dot{M}_{\rm shut}$ is a more convenient quantity (or independent variable) than time to refer to, because, first, the initial conditions of planet formation as well as the age of the system are often poorly constrained. And, second, accretion rates can be inferred more easily from observations than age. Hence, planetary accretion rates allow us to consider planet (and moon) formation at comparable stages, independent of the initial conditions.

 \begin{figure}[tH!]
   \centering
   \scalebox{0.62}{\includegraphics{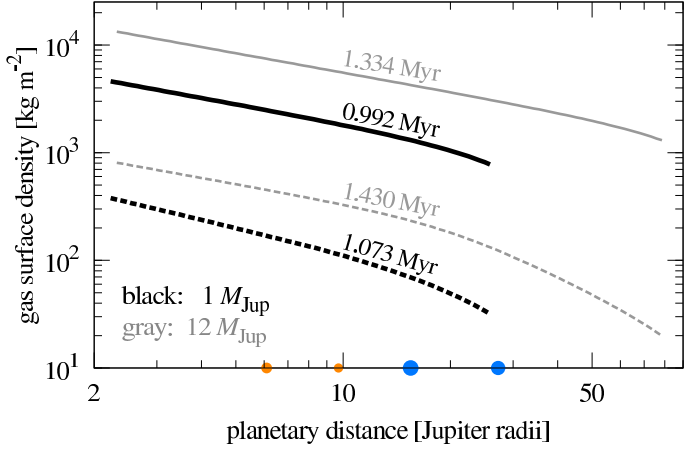}}
   \caption{Surface densities around the four test planets shown in Figs.~\ref{fig:T_shots_1} and \ref{fig:T_shots_12}. Black lines refer to the Jupter-mass planets, gray lines to the $12\,M_{\rm Jup}$-mass super-Jovian. Note that the disk of the Jupiter twin is much smaller than that of the super-Jovian and that both surface density distributions decrease with time. The radial positions of the Galilean moons are indicated at the bottom of the figure.}
   \label{fig:Sigma_shots}
 \end{figure}

Variations of the stellar distance will affect both the size of the circumplanetary accretion disk \citep{2008ApJ...685.1220M} and the stellar heating term in our model \citep{2014SoSyR..48...62M}. Disks around close-in planets are usually smaller (though not necessarily less massive) than disks around planets in wide stellar orbits owing to their smaller Hill spheres and their lower average specific angular momentum. Moreover, substantial stellar illumination will prevent these disks from having H$_2$O ice lines in the vicinity of the star. The temperature of the circumstellar accretion disk, in which the planet is embedded, is calculated under the assumption that the disk is transparent to stellar irradiation \citep{1981PThPS..70...35H} and that the stellar luminosity equals that of the contemporary Sun. In this model, the circumstellar H$_2$O ice line is located at 2.7\,AU from the star. We will revisit the faint young Sun \citep{1972Sci...177...52S} as well as other parameterizations of the circumstellar disk in an upcoming paper \citep[][in prep.]{HMP2015}.

In Figs.~\ref{fig:T_shots_1} and \ref{fig:T_shots_12} we show an application of our disk model to a Jupiter-mass and a $12\,M_{\rm Jup}$ planet at 5.2\,AU around a Sun-like star, respectively. All panels present the disk surface temperatures ($T_{\rm s}$, solid gray lines) and disk midplane temperatures ($T_{\rm m}$, solid black lines) as well as the contributions to $T_{\rm s}$ from viscous heating ($T_{\rm v}$, black crosses), planetary illumination ($T_{\rm p}$, solid black lines), accretion onto the disk ($T_{\rm a}$, gray dashed lines), and heating from the circumstellar accretion disk, or ``nebula'' ($T_{\rm n}$, gray dotted lines). Any of these contributions to the disk surface temperature is computed assuming that all other contributions are zero. In other words, $T_{\rm v}$, $T_{\rm p}$, $T_{\rm a}$, and $T_{\rm n}$ depict the temperature of the disk photosphere in the hypothetical case that viscous heating, planetary illumination, accretional heating onto the disk, or the stellar illumination were the single energy source, respectively.\footnote{As an example, $T_{\rm v}$ is calculated setting $F_{\rm acc}~=~F_{\rm p}~=~T_{\rm neb}~=0$ in Eq.~(13) of \citet{2014arXiv1410.5802H}.} The different slopes of these curves, in particular of $T_{\rm v}$ and $T_{\rm p}$, lead to the appearance of heat transitions (not shown), which traverse the disk on relatively short time scales \citep{2014arXiv1410.5802H}. Also note that in all the simulations shown, the disk midplane is warmer than the disk surface. Only when accretion drops to about $10\,M_{\rm Gan}\,{\rm Myr}^{-1}$ in panels (b) do $T_{\rm s}$ and $T_{\rm m}$ become comparable throughout the disk, both around the Jupiter- and the super-Jupiter-mass planet.

 \begin{figure*}[t]
   \centering
   \scalebox{0.61}{\includegraphics{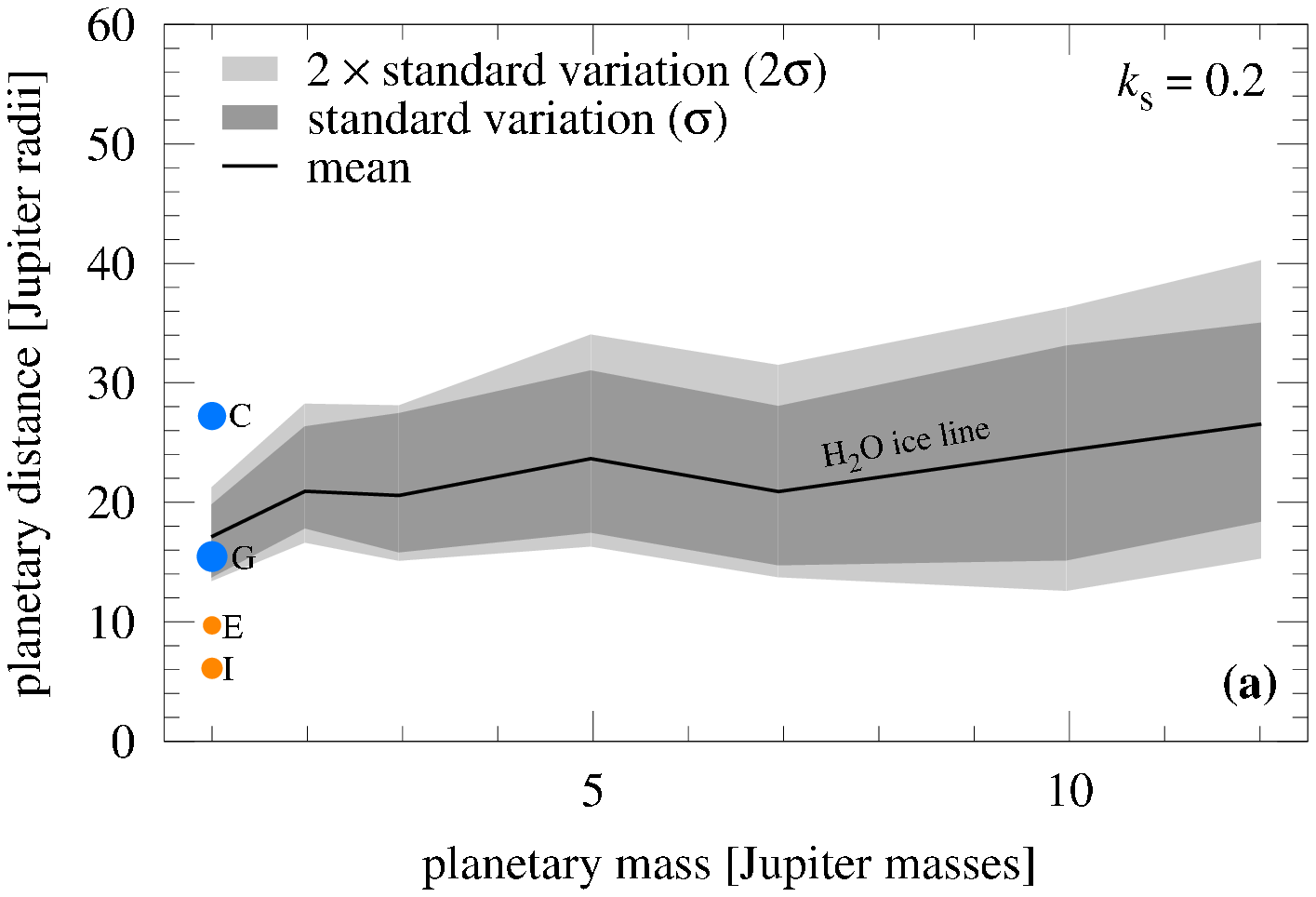}}
   \hspace{0.66cm}
   \scalebox{0.61}{\includegraphics{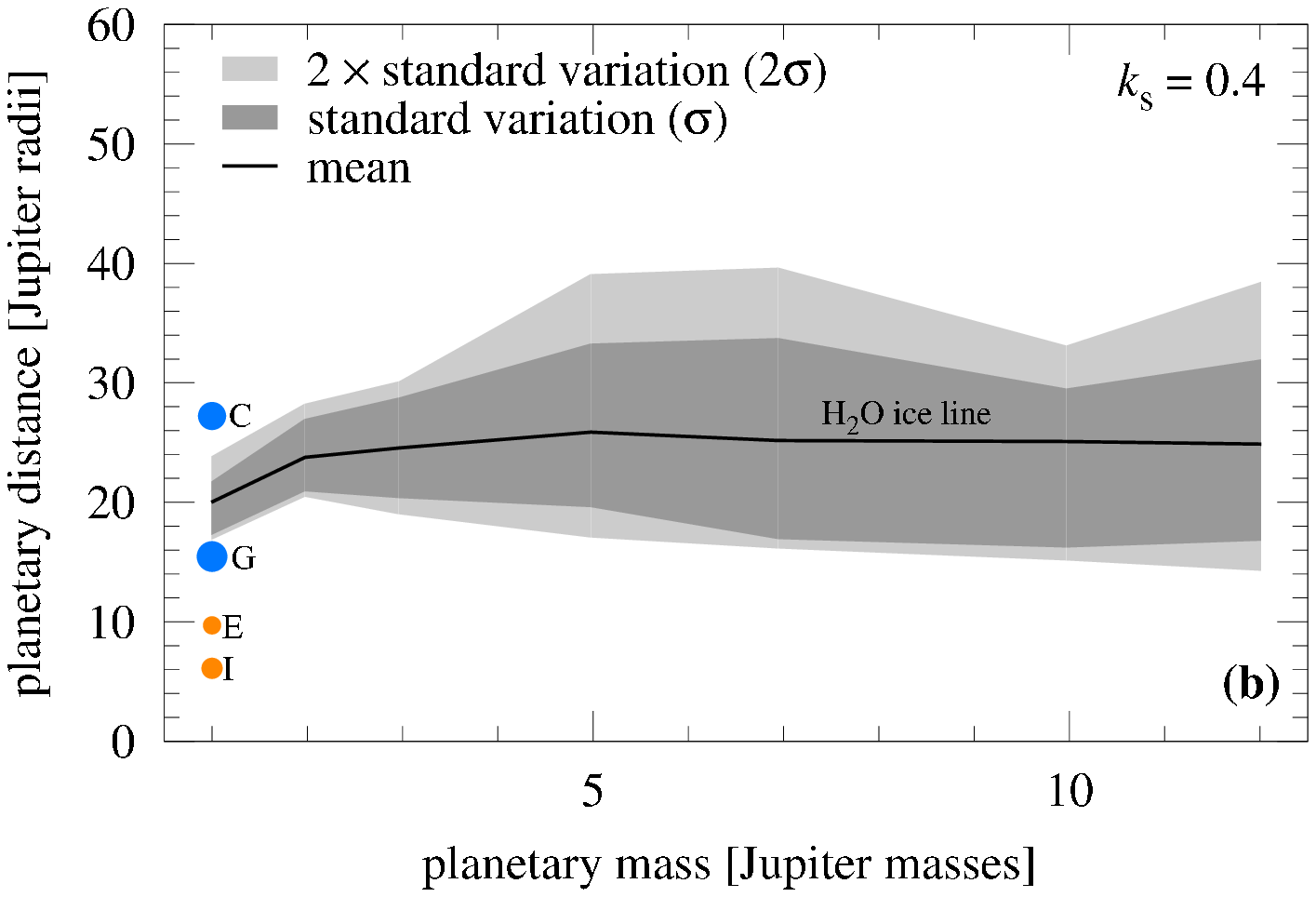}}
   \caption{Distance of the circumplanetary H$_2$O ice lines as a function of planetary mass. A distance of 5.2\,AU to a Sun-like star is assumed. The shaded area indicates the 1$\sigma$ scatter of our simulations based on the posterior distribution of the disk Planck mean opacity ($\kappa_\mathrm{P}$) and the shutdown accretion rate for moon formation ($\dot{M}_\mathrm{shut}$). The labeled circles at $1\,M_\mathrm{Jup}$ denote the orbital positions of Jupiter's moons Io, Europa, Ganymede, and Callisto. Orange indicates rocky composition, blue represents H$_2$O-rich composition. Circle sizes scale with the moons' radii. \textbf{(a)}: Disk reflectivity ($k_\mathrm{s}$) is set to $0.2$. \textbf{(b)}: Same model parameterization but now with disk reflectivity $k_\mathrm{s}=0.4$.}
   \label{fig:stats}
 \end{figure*}

 \begin{figure*}[t]
   \centering
   \scalebox{0.6}{\includegraphics{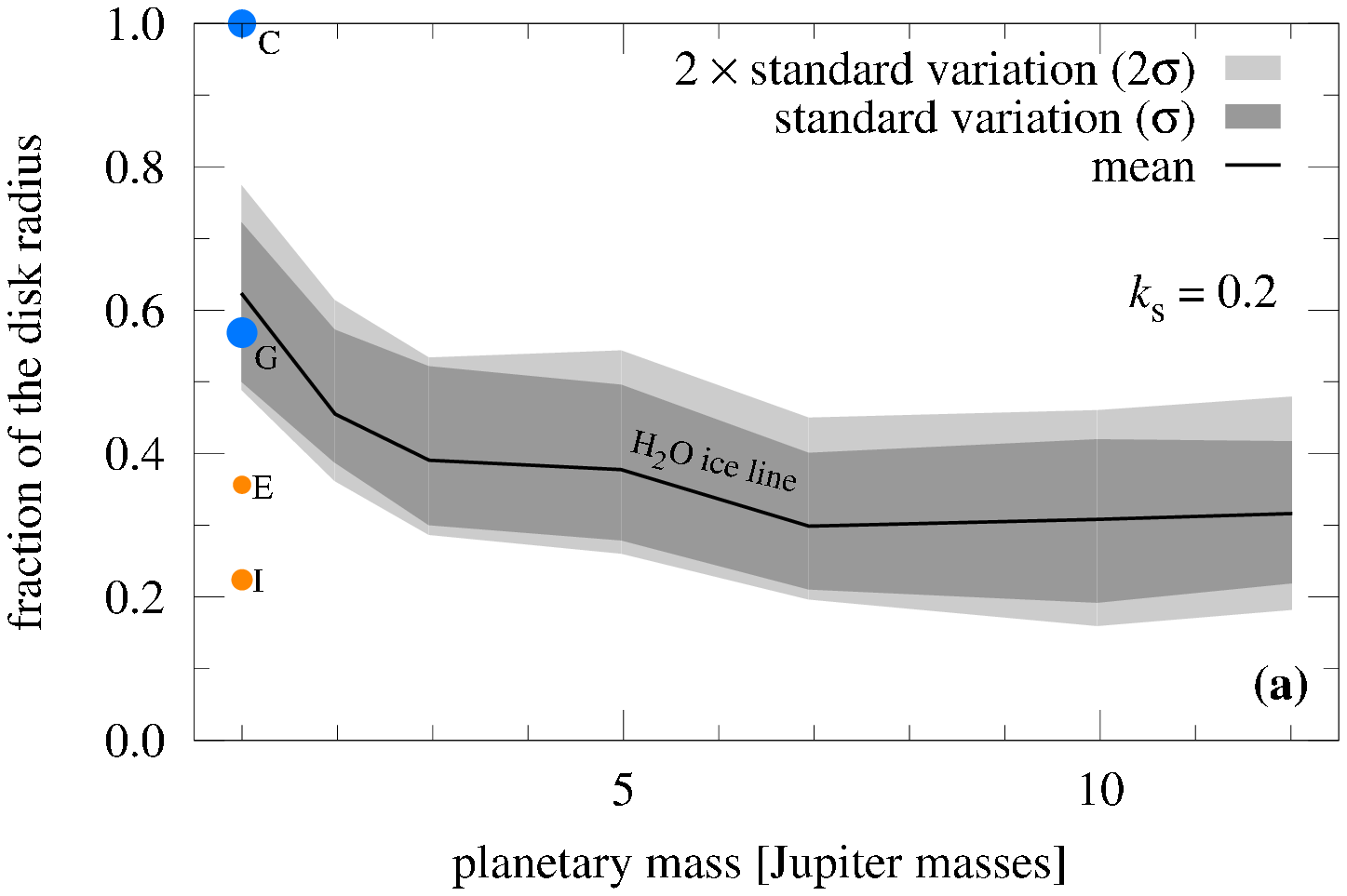}}
   \hspace{0.66cm}
   \scalebox{0.6}{\includegraphics{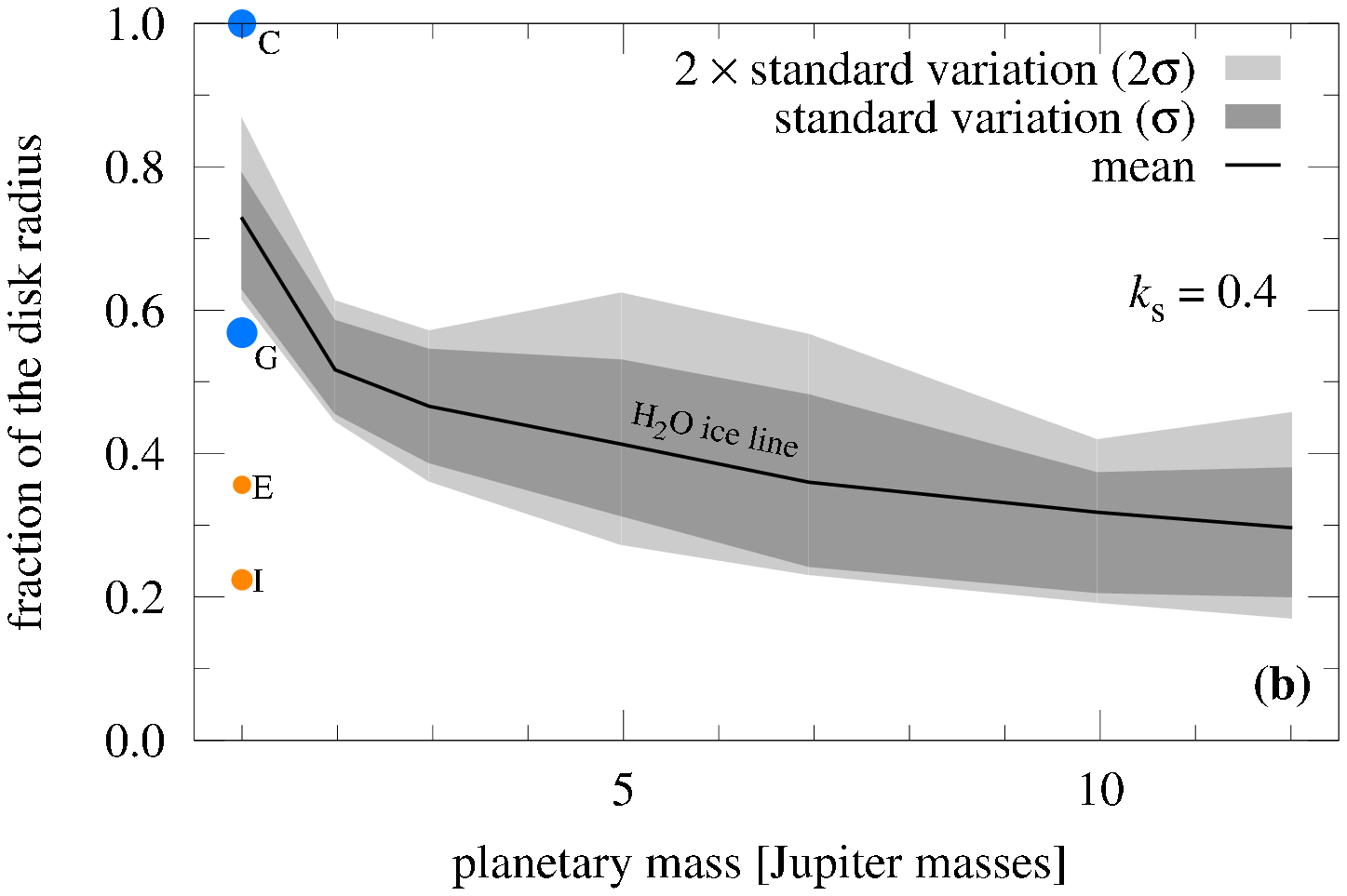}}
   \caption{Similar to Fig.~\ref{fig:stats}, but now in units of fractional disk radius (see ordinate). \textbf{(a)}: Disk reflectivity ($k_\mathrm{s}$) is set to $0.2$. \textbf{(b)}: Same model parameterization but now with disk reflectivity $k_\mathrm{s}=0.4$. The negative slope in both panels indicates that the most massive giant planets have larger fractions of their disks fed with water ice and, consequently, relatively more space and material to form big moons.}
   \label{fig:statsfrac}
 \end{figure*}

Panels (a) in Figs.~\ref{fig:T_shots_1} and \ref{fig:T_shots_12} are chosen at the time at which the H$_2$O ice line first appears at the outer edge of the disk, while panels (b) illustrate the temperature distribution at the time when $\dot{M}=10\,M_{\rm Gan}$, with $M_{\rm Gan}$ as Ganymede's mass. During this epoch, the H$_2$O ice line around the Jupiter-mass planet is safely between the orbital radii of Europa and Ganymede (see labeled arrow in Fig.~\ref{fig:T_shots_1}b), which suggests that this corresponds to the shutdown phase of moon formation. The colored symbols in each panel depict the rocky (orange) or icy (blue) composition of the Galilean moons. Symbol sizes scale with the actual moon radii. In these simulations, the Planck mean opacity ($\kappa_{\rm P}$) has been fixed at a fiducial value of $10^{-2}\,{\rm m}^2\,{\rm kg}^{-1}$ throughout the disk and the disk absorptivity is assumed to be $k_{\rm s}=0.2$.

A comparison of Figs.~\ref{fig:T_shots_1}(a) and \ref{fig:T_shots_12}(a) shows that the centrifugal radius of the accretion disk around the Jovian planet ($r_{\rm c}\approx27\,R_{\rm Jup}$) is substantially smaller than the disk radius around the super-Jovian test planet ($r_{\rm c}\approx80\,R_{\rm Jup}$) -- note the different distance ranges shown in the figures! Moreover, the midplanes and surfaces in the inner regions of the super-Jovian accretion disks are substantially hotter at any given distance around the super-Jupiter. Note also that planetary illumination is the main energy source around the Jupiter-mass planet in Figs.~\ref{fig:T_shots_1}(a) and (b), while viscous heating plays an important role during the appearance of the H$_2$O ice line around the super-Jovian planet in Fig.~\ref{fig:T_shots_12}(a). In the final stages of accretion onto the $12\,M_{\rm Jup}$ planet, shown in Fig.~\ref{fig:T_shots_12}(b), viscous heating and planetary illumination are comparable throughout the disk.

In Fig.~\ref{fig:Sigma_shots} we present the radial distributions of the gas surface densities around our Jovian and super-Jovian test planets. While solid lines refer to panels (a) in Figs.~\ref{fig:T_shots_1} and \ref{fig:T_shots_12}, dashed lines refer to panels (b), respectively. In particular, solid lines refer to that instant in time when the H$_2$O ice lines first appear at the outer edges of the accretion disks around those two planets, whereas the dashed lines depict the accretion phase when $\dot{M}~=~10\,M_{\rm Gan}\,{\rm Myr}^{-1}$, which is the phase when the H$_2$O ice line around our Jupiter-mass test planet is in good agreement with the compositional gradient observed in the Galilean system \citep{2014arXiv1410.5802H}. At any given planetary distance in these particular states, the super-Jovian planet has a gas surface density that is higher by about a factor of five compared to the Jupiter-mass planet. Moreover, we find that the H$_2$O ice lines in both cases need about $10^5$\,yr to move radially from the outer disk edges to their final positions (see labels). Our values for $\Sigma_{\rm g}$ are similar to those presented by \citet{2014SoSyR..48...62M} in their Fig.~4.

Similar to the procedure applied in \citet{2014arXiv1410.5802H}, we performed a suite of simulations for planets with masses between 1 and $12\,M_{\rm Jup}$ at 5.2\,AU from a Sun-like star, where $\dot{M}_{\rm shut}$ and $\kappa_{\rm P}$ were randomly drawn from a Gaussian probability distribution. For $\log_{10}(\kappa_{\rm P}/[{\rm m}^2\,{\rm kg}^{-1}])$ we took a mean value of $-2$ with a standard variation of 1 \citep[see][and references therein]{1997ApJ...486..372B}, and for $\log_{10}(\dot{M}_{\rm shut}/[M_{\rm Gan}\,{\rm Myr}^{-1}])$ we assumed a mean value of 1 with a standard variation of 1, which nicely reproduced the compositional H$_2$O gradient in the Galilean moons \citep{2014arXiv1410.5802H}. As an extension of our previous simulations, we here consider various disk absorptivities ($k_{\rm s}=0.2$ and 0.4) and examine the total mass of solids in the accretion disks, both as a function of time and as a function of stellar distance. Above all, we study the disappearance of the circumplanetary H$_2$O ice line in the vicinity of the star.

\section{H$_2$O ice lines around planets of different masses}
\label{sec:Mp}

Figures~\ref{fig:T_shots_1}(a) and (b) show that the temperature distribution in the late accretion disks around Jupiter-mass planets is determined mostly by the planetary illumination rather than viscous heating. This indicates a key difference between moon formation in circumplanetary accretion disks and planet formation in circumstellar disks, where the positions of ice lines have been shown to depend mostly on viscous heating rather than stellar illumination \citep{2011Icar..212..416M,2011MNRAS.417.1236H}. In this context, Figs.~\ref{fig:T_shots_12}(a) and (b) depict an interesting intermediate case in terms of the mass of the central object, where both viscous heating and illumination around a $12\,M_{\rm Jup}$ planet show comparable contributions towards the final stages of accretion in panel (b). Also note that in Fig.~\ref{fig:T_shots_12}(b) the temperature distribution in the disk outskirts is determined by the background temperature provided by stellar illumination. This suggests that the extended disks around super-Jovian exoplanets in wide circumstellar orbits (beyond about 10\,AU) might also feature CO and other ice lines due to the even weaker stellar illumination.

Figures~\ref{fig:stats}(a) and (b) display the radial positions of the circumplanetary H$_2$O ice lines around a range of planets with masses between 1 and $12\,M_{\rm Jup}$ in the final stages of accretion, where we randomized $\kappa_{\rm P}$ and $\dot{M}_{\rm shut}$ as described above, and the stellar distance is 5.2\,AU for all planets considered. Panel (a) shows the same data as Fig.~4 in \citet{2014arXiv1410.5802H}, while panel (b) assumes $k_{\rm s}=0.4$. The larger absorptivity in panel (b) pushes the H$_2$O ice lines slightly away from the accreting planets compared to panel (a) for all planets except for the $12\,M_{\rm Jup}$ planet, which we ascribe to an insignificant statistical fluctuation. As a key result of our new simulations for $k_{\rm s}=0.4$, and in agreement with our previous study, the H$_2$O ice line is between about 15 and $30\,R_{\rm Jup}$ for all super-Jovian planets and almost independent of $M_{\rm p}$. The effect of changing $k_{\rm s}$ by a factor of two is moderate, pushing the H$_2$O ice line outward by only a few $R_{\rm Jup}$ on average.

Figure~\ref{fig:statsfrac} presents a different visualization of these results, again with panel (a) assuming $k_{\rm s}=0.2$ and panel (b) assuming $k_{\rm s}=0.4$. Now the ordinate gives the circumplanetary distance in units of the fractional disk radius, and so the H$_2$O ice lines are located closer to the planet with increasing $M_{\rm p}$. This is a consequence of the larger disk sizes of the more massive planets. In other words, larger giant planets have larger fractions of the accretion disks beyond the H$_2$O ice lines. Naturally, this means that these super-Jovians should form the most massive, icy moons.

\begin{figure}[tH!]
   \centering
   \scalebox{0.61}{\includegraphics{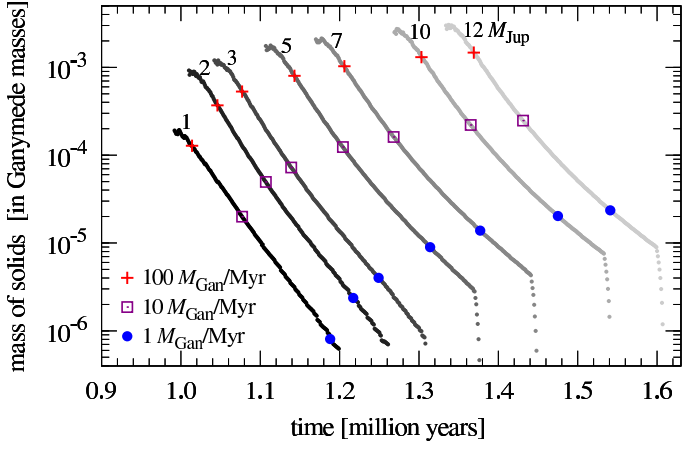}}
   \caption{Evolution of the total mass of solids ($M_{\rm sld}$) in circumplanetary accretion disks of several test planets at 5.2\,AU from a Sun-like star. Colored dots along the mass evolution tracks indicate accretion rates ($\dot{M}$). If moon formation around all of these planets shuts down at comparable disk accretion rates ($\dot{M}_{\rm shut}$), then $M_{\rm sld}(\dot{M}_{\rm shut})\,{\propto}\,M_{\rm p}$ \citep[see also Fig.~5 in][]{2014arXiv1410.5802H}. Compared to a Jupiter-mass planet, a $12\,M_{\rm Jup}$ planet would then have 12 times the amount of solids available for moon formation.}
   \label{fig:MsInst_t}
 \end{figure}

In Fig.~\ref{fig:MsInst_t}, we therefore analyze the evolution of $M_{\rm sld}$ (in units of $M_{\rm Gan}$) around these planets as a function of time. Indicated values for $\dot{M}$ along these tracks help to visualize $M_{\rm sld}$ as a function of accretion rates (see crosses, squares, and circles), because time on the abscissa delivers an incomplete picture of the mass evolution due to the different timescales of these disks. We find that disks around lower-mass super-Jovians contain less solid mass at any given accretion rate than disks around higher-mass super-Jovians. Solids also occur earlier in time as the appearance of the disk itself is regulated by the shrinking of the initially very large planet in our models: lower-mass giant planets contract earlier \citep{2013A&A...558A.113M}. Assuming that moon formation shuts down at similar accretion rates around any of the simulated planets, the example accretion rates (100, 10, $1\,M_{\rm Gan}\,{\rm Myr}^{-1}$) indicate an increase of the mass of solids as a function of $M_{\rm p}$. For a given $\dot{M}_{\rm shut}$, this scaling is $M_{\rm sld}~{\propto}~M_{\rm p}$ \citep[see also Fig.~5 in][]{2014arXiv1410.5802H}, which is in agreement with the scaling relation for the total moon mass around giant planets found by \citet{2006Natur.441..834C}.

\section{H$_2$O ice lines and total mass of solids around planets at various stellar distances}
\label{sec:a}

Having examined the sensitivity of our results to $M_{\rm p}$ and to the disk's radiative properties, we now turn to the question of what exomoon systems are like around super-Jovian exoplanets at very different locations in their disks than our own Jupiter at 5.2\,AU. Here, we shall encounter some significant surprises.

\begin{figure*}[t]
   \centering
   \scalebox{0.625}{\includegraphics{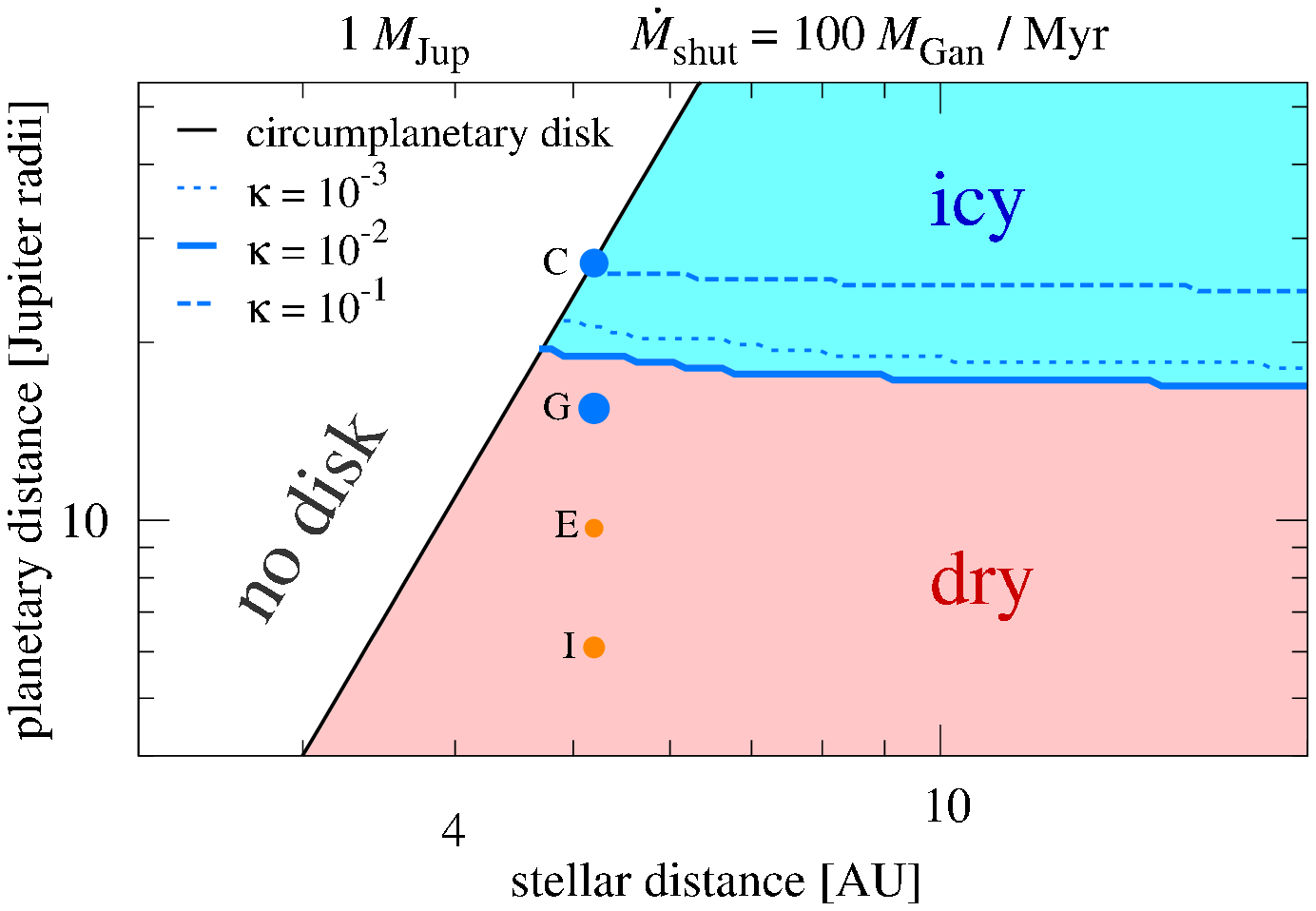}}
   \hspace{0.66cm}
   \scalebox{0.625}{\includegraphics{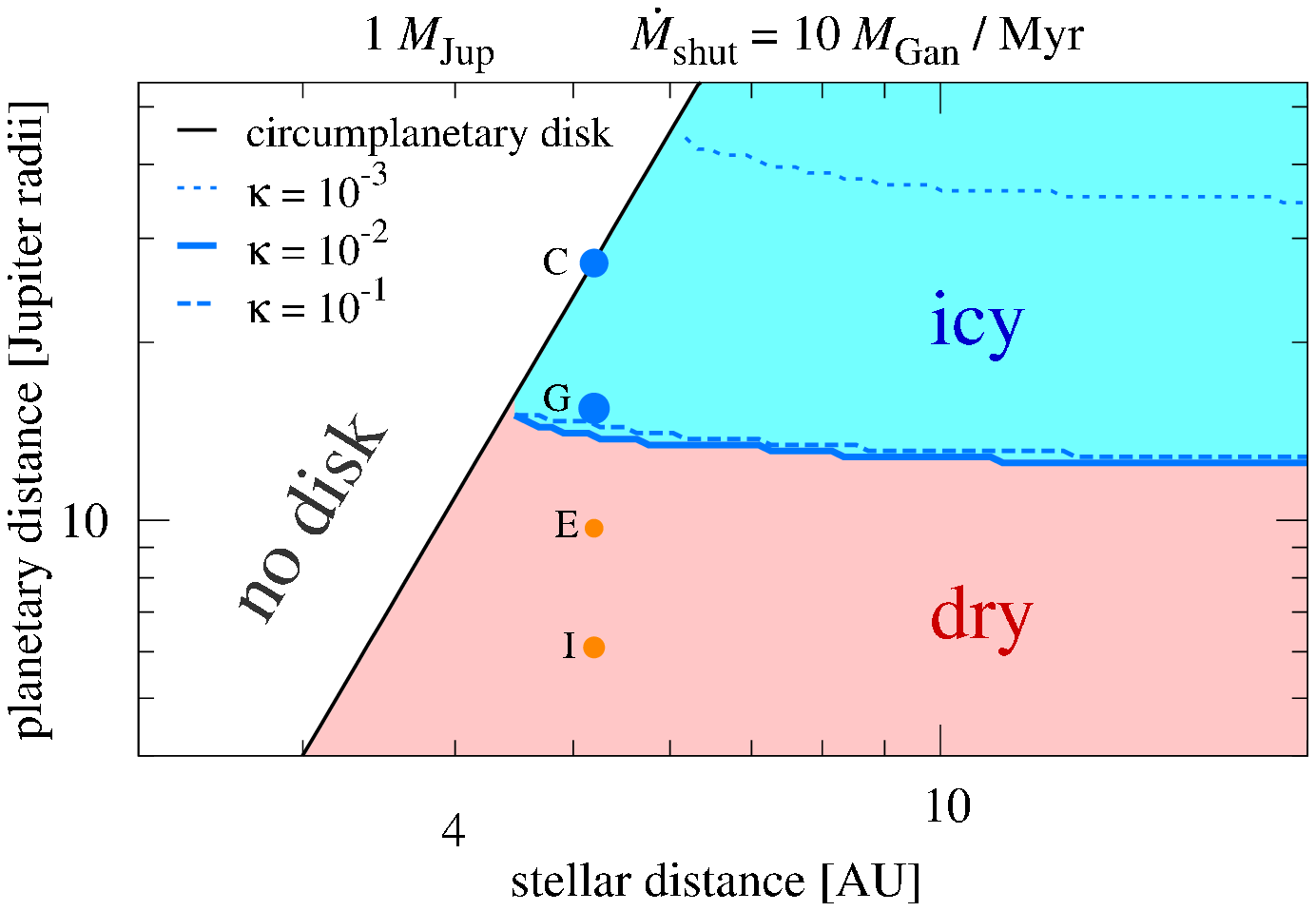}}
   \caption{Distances of the circumplanetary H$_2$O ice lines around a Jupiter-like planet (ordinate) as a function of distance to a Sun-like star (abscissa) at the time of moon formation shutdown. Each panel assumes a different shutdown formation rate (see panel titles). Three Planck opacities through the circumplanetary disk are tested in each panel (in units of ${\rm m}^2\,{\rm kg}^{-1}$, see panel legends). The circumplanetary orbits of the Galilean satellites are represented by symbols as in Fig.~\ref{fig:T_shots_1}. The black solid line shows the disk's centrifugal radius \citep{2008ApJ...685.1220M}. \textbf{(a)} At  $\dot{M}_{\rm shut} = 100\,M_{\rm Gan}/{\rm Myr}$ the H$_2$O ice lines at $\approx5$\,AU are about $5\,R_\mathrm{Jup}$ beyond Ganymede's current orbit. \textbf{(b)} At  $\dot{M}_{\rm shut} = 10\,M_{\rm Gan}/{\rm Myr}$ values of $10^{-2}\,{\rm m}^2\,{\rm kg}^{-1}\leq\kappa\leq10^{-1}\,{\rm m}^2\,{\rm kg}^{-1}$ place the H$_2$O ice line slightly inside the current orbit of Ganymede and thereby seem most plausible.}
   \label{fig:distance_1Mjup}
 \end{figure*}

Figure~\ref{fig:distance_1Mjup}(a) shows the final circumplanetary distance of the H$_2$O ice line around a Jupiter-mass planet between 2 and 20\,AU from the star,\footnote{We also simulated planets as close as 0.2\,AU to the star, but their accretion disks are nominally smaller than the planetary radius, indicating a departure of our model from reality. Anyways, since the hypothetical disks around these planets do not harbor H$_2$O ice lines and moon formation in the stellar vicinity is hard in the first place \citep{2002ApJ...575.1087B,2010ApJ...719L.145N}, we limit Fig.~\ref{fig:MsInst} to 2\,AU.} assuming a shutdown accretion rate of $100\,M_{\rm Gan}\,{\rm Myr}^{-1}$. Different styles of the blue lines correspond to different disk opacities (see legend), while the solid black line indicates the radius of the circumplanetary accretion disk, following \citet{2008ApJ...685.1220M}. Circles indicate the radial distances of the Galilean moons around Jupiter, at 5.2\,AU from the star. In this set of simulations, the H$_2$O ice lines always ends up between the orbits of Ganymede and Callisto, which is not in agreement with the observed H$_2$O compositional gradient in the Galilean system. As stellar illumination decreases at larger distances while all other heating terms are constant for the given accretion rate, the ice lines move towards the planet at greater distances.

Most importantly, however, we find that Jovian planets closer than about 4.8\,AU do not have an H$_2$O ice line in the first place. Hence, if the large population of Jupiter-mass planets around 1\,AU (see Fig.~\ref{fig:exoplanets}) formed in situ and without substantial inward migration from the outer regions, then these giant planets should not have had the capacity to form giant, icy moons, that is, scaled-up versions of Ganymede or Callisto. These giant moons with masses up to that of Mars  \citep[suggested by][]{2006Natur.441..834C,2014AsBio..14..798H,2014arXiv1410.5802H}, may only be present if they have completed their own water-rich formation beyond about 4.8\,AU from their star, before they migrated to their current circumstellar orbits together with their host planets.

 \begin{figure*}[t]
   \centering
   \scalebox{0.625}{\includegraphics{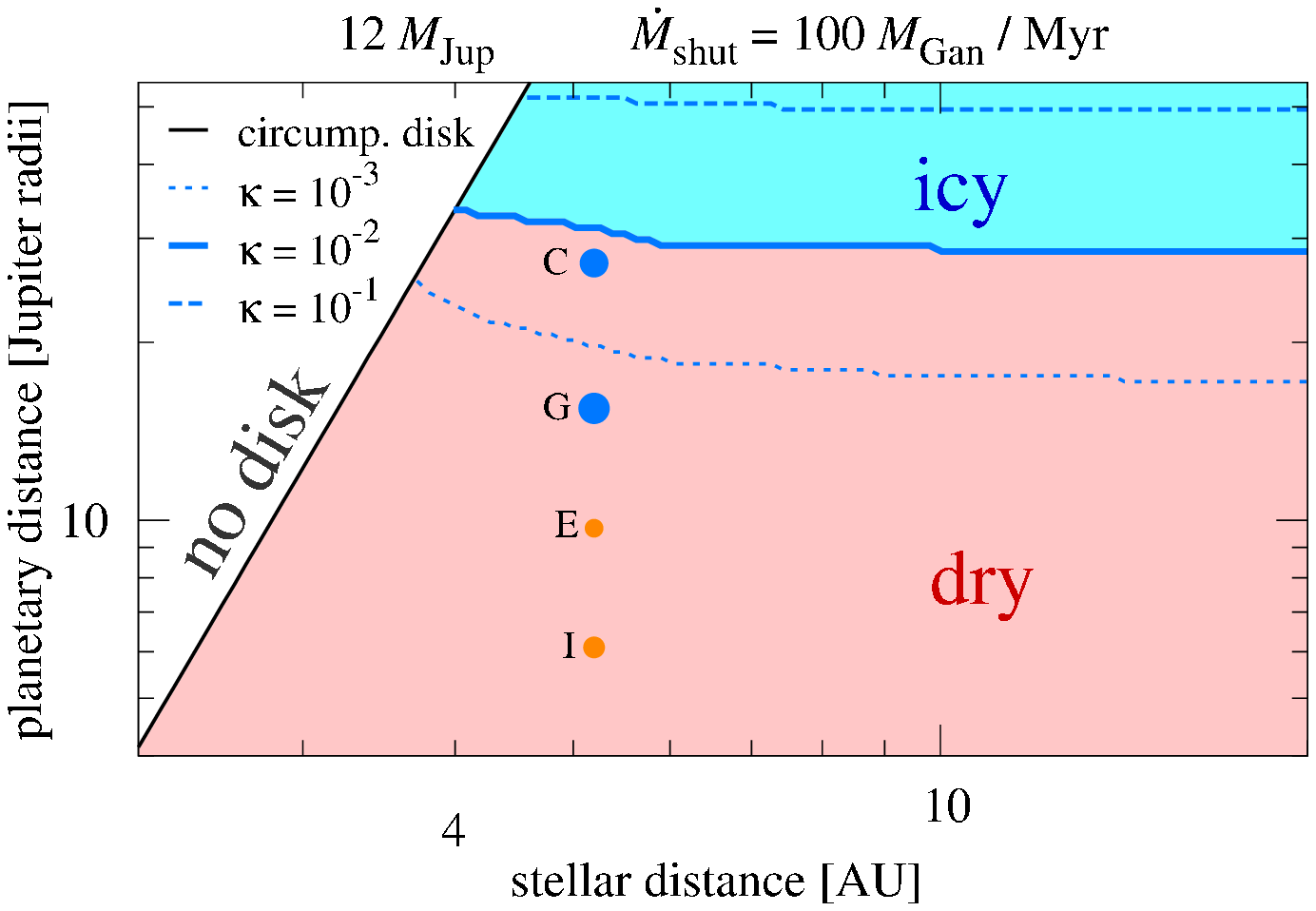}}
   \hspace{0.66cm}
   \scalebox{0.625}{\includegraphics{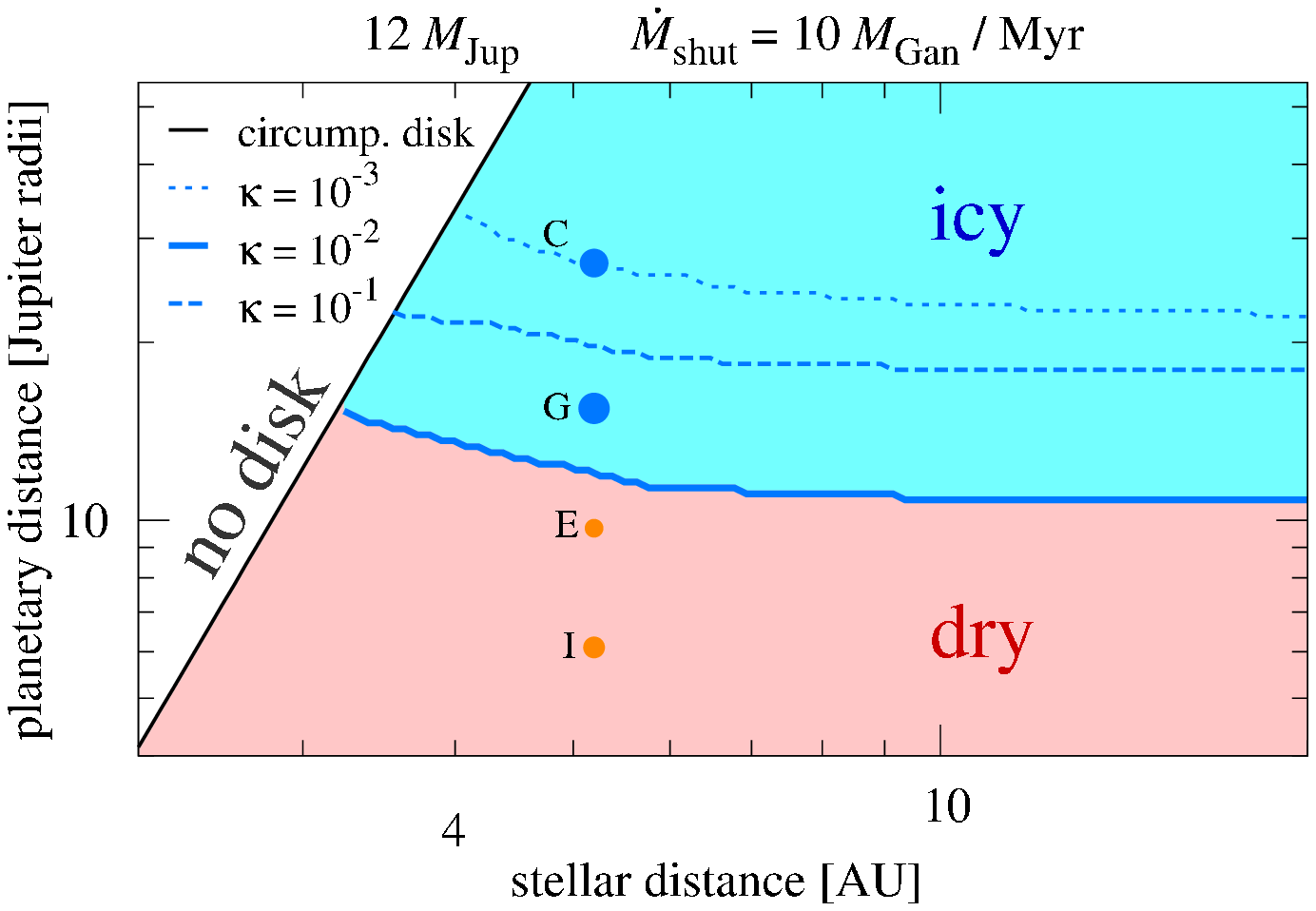}}
   \caption{Same as Fig.~\ref{fig:distance_1Mjup}, but now for a $12\,M_{\rm Jup}$ mass planet.}
   \label{fig:distance_12Mjup}
 \end{figure*}

 \begin{figure*}[t]
   \centering
   \scalebox{0.61}{\includegraphics{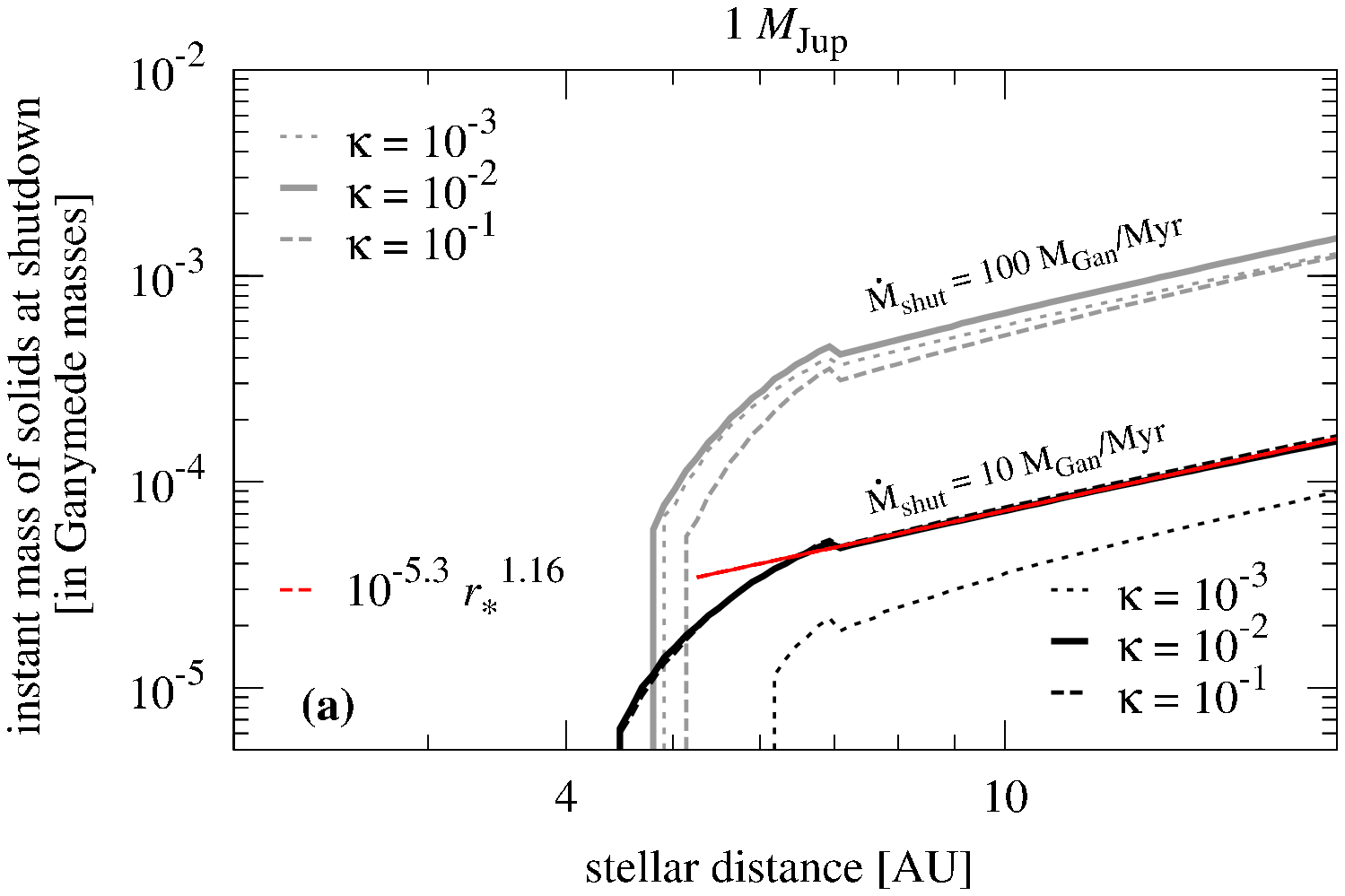}}
   \hspace{0.3cm}
   \scalebox{0.61}{\includegraphics{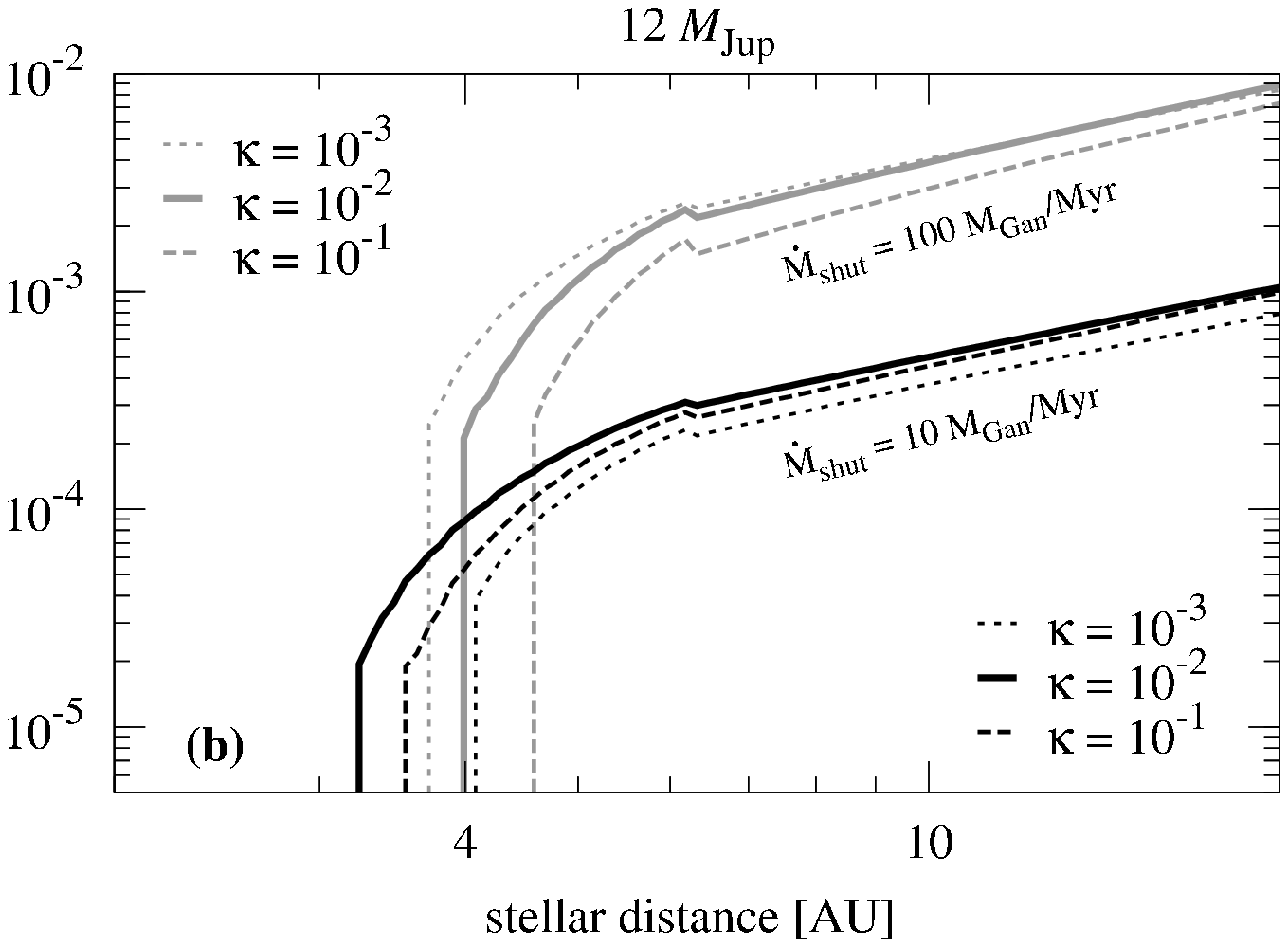}}
   \caption{Total instantaneous mass of solids in circumplanetary accretion disks in units of Ganymede masses and as a function of stellar distance. Labels indicate two different values for $\dot{M}_{\rm shut}$, and different line types refer to different disk opacities (see legend). Black dashed lines in both panels indicate our fiducial reference disk. \textbf{(a)}: The $1\,M_{\rm Jup}$ test planet starts to have increasingly large amounts of solids only beyond 4.5\,AU, depending on the accretion rate and disk opacity. The red dashed line indicates our fit as per Equation~\eqref{eq:M_sld}. \textbf{(b)}: The $12\,M_{\rm Jup}$ test planet starts to contain substantial amounts of solids as close as about 3.1\,AU to the star, owed to its larger disk.}
   \label{fig:MsInst}
 \end{figure*}

In Fig.~\ref{fig:distance_1Mjup}(b), the planetary accretion rate has dropped by a factor of ten, and the ice lines have shifted. For $\kappa_{\rm P}~=~10^{-2}\,{\rm m}^2\,{\rm kg}^{-1}$ and $\kappa_{\rm P}~=~10^{-1}\,{\rm m}^2\,{\rm kg}^{-1}$ they moved towards the planet. But for  $\kappa_{\rm P}~=~10^{-3}\,{\rm m}^2\,{\rm kg}^{-1}$ they moved outward. The former two rates actually place the H$_2$O ice line around Jupiter at almost exactly the orbit of Ganymede, which is in better agreement with observations. In these simulations, Jovian planets closer than about 4.5\,AU do not have a circumplanetary H$_2$O ice line.

In Fig.~\ref{fig:distance_12Mjup} we vary the stellar distance of a $12\,M_{\rm Jup}$ planet. First, note that the disk (black solid line) is larger at any given stellar separation than in Fig.~\ref{fig:distance_1Mjup}. Second, note that the H$_2$O ice line for a given $\dot{M}_{\rm shut}$ in Figs.~\ref{fig:distance_12Mjup}(a) and (b), respectively, is further out than in the former case of a Jupiter-mass planet. This is due to both increased viscous heating and illumination from the planet for the super-Jovian object. Nevertheless, the much larger disk radius overcompensates for this effect and, hence, all these simulations suggests that accretion disks around the most massive planets can have H$_2$O ice lines if the planet is not closer than about 3.8\,AU (panel a) to 3.1\,AU (panel b) to a Sun-like star. The latter value refers to a shutdown accretion rate of $10\,M_{\rm Gan}\,{\rm Myr}^{-1}$, which is in good agreement with the water ice distribution in the Galilean moon system. Assuming that moon formation stops at comparable accretion rates around super-Jovian planets, we consider a critical stellar distance of about 3\,AU a plausible estimate for the critical stellar distance of a $12\,M_{\rm Jup}$ accreting planet to show a circumplanetary H$_2$O ice line.

Figure~\ref{fig:MsInst} presents the total instantaneous mass of solids in circumplanetary accretion disk as a function of stellar distance for two different shutdown accretion rates and three different disk Planck opacities. The calculation of $M_{\rm sld}$ follows Eq.~(17) in \citet{2014arXiv1410.5802H}, that is, we integrate the surface density of solids between the inner disk truncation radius and the outer centrifugal radius. Panel (a) for a Jupiter-mass planet demonstrates that indeed the amount of solids in its late accretion disk is negligible within about 4.5\,AU from a Sun-like star. This gives us crucial insights into Jupiter's migration history within the Grand Tack framework \citep{2011Natur.475..206W}, which we will present in a forthcoming paper \citep[][in prep.]{HMP2015}. Moreover, both panels show that the effect of different disk opacities on $M_{\rm sld}$ is small for a given $M_{\rm p}$ and $\dot{M}_{\rm shut}$.

Intriguingly, for any given disk parameterization (see legend) a circumjovian disk at 5.2\,AU from a Sun-like star seems to harbor a relatively small amount of solids compared to disks around planets at larger stellar distances. The red dashed line indicates a best fit exponential function to the simulations of our fiducial disk with $\dot{M}_{\rm shut}=10\,M_{\rm Gan}\,{\rm Myr}^{-1}$ and $\kappa_{\rm P}=10^{-2}\,{\rm m}^2\,{\rm kg}^{-1}$ beyond 5.2\,AU, as an example. We choose this disk parameterization as it yields the best agreement with the radial location of the icy Galilean satellites \citep{2014arXiv1410.5802H}. It scales as

\begin{equation}\label{eq:M_sld}
M_{\rm sld} = 10^{-5.3}\,M_{\rm Gan} \ \times \ \left( \frac{r_\star}{{\rm AU}} \right)^{1.16} \ \ , 
\end{equation}

\noindent
where $r_\star$ is the stellar distance. Assuming that moon formation stops at similar mass accretion rates around any super-Jovian planet and taking into account the previously known scaling of the total moon masses ($M_{\rm T}$) with $M_{\rm p}$, we deduce a more general estimate for the total moon mass:

\begin{equation}\label{eq:M_T}
M_{\rm T} = M_{\rm GM} \ \times \left( \frac{M_{\rm p}}{M_{\rm Jup}} \right) \ \times \ \left( \frac{r_\star}{5.2\,{\rm AU}} \right)^{1.16} \ , \ (r_\star \geq 5.2\,{\rm AU}) \ , 
\end{equation}

\noindent
where $M_{\rm GM}=2.65\,M_{\rm Gan}$ is the total mass of the Galilean moons. As an example, a $10\,M_{\rm Jup}$ planet forming at 5.2\,AU around a Sun-like star should have a moon system with a total mass of about $10\,M_{\rm GM}$ or 6 times the mass of Mars. At a Saturn-like stellar distance of 9.6\,AU, $M_{\rm T}$ would be doubled. Simulations by \citet{2014AsBio..14..798H} show that this mass will be distributed over three to six moons in about 90\,\% of the cases. Hence, if the most massive super-Jovian planets formed moon systems before they migrated to about 1\,AU, where we observe them today (see Fig.~\ref{fig:exoplanets}), then Mars-mass moons in the stellar habitable zones might be very abundant.

\section{Shutdown accretion rates and loss of moons}
\label{sec:general}

\citet{2002AJ....124.3404C} argued that accretion rates of $2~\times~10^{-7}\,M_\mathrm{Jup}\,\mathrm{yr}^{-1}$ (or about $2.6~\times10^{4}\,M_{\rm Gan}\,\mathrm{Myr}^{-1}$) best reproduce the disk conditions in which the Galilean system formed. Based on the condition that the H$_2$O ice line needs to be between the orbits of Europa and Ganymede at the final stages of accretion, our calculations predict a shutdown accretion rate that is considerably lower, closer to $10\,M_{\rm Gan}\,\mathrm{Myr}^{-1}$ (see Fig.~\ref{fig:distance_1Mjup}). The difference in these results is mainly owed to two facts. First, \citet{2002AJ....124.3404C} only considered viscous heating and planetary illumination. Our additional heating terms (accretion onto the disk and stellar illumination) contribute additional heat, which imply smaller accretion rates to let the H$_2$O ice lines move close enough to the Jupiter-like planet. Second, the parameterization of planetary illumination in the \citet{2002AJ....124.3404C} model is different from ours. While \citet{2002AJ....124.3404C} assume an $r^{-3/4}$ dependence of the midplane temperature from the planet ($r$ being the planetary radial distance), we do not apply any predescribed $r$-dependence. In particular, $T_{\rm m}(r)$ cannot be described properly by a simple polynomial due to the different slopes of the various heat sources as a function of planetary distance (see the black solid lines in Figs.~\ref{fig:T_shots_1} and \ref{fig:T_shots_12}).

While our estimates of $\dot{M}_{\rm shut}$ are about three orders of magnitude lower than the values proposed by \citet{2002AJ....124.3404C}, they are also one to two orders of magnitude lower than the values suggested by \citet{2005A&A...439.1205A}. Their moon formation model is similar to the so-called ``gas-starved'' disk model applied by \citet{2002AJ....124.3404C}. Accretion rates in their model were not derived from planet formation simulations (as in our case) but calculated using an analytical fit to previous simulations. Again, the additional energy terms in our model are one reason for the lower final accretion rates that we require in order to have the H$_2$O ice line interior to the orbit of Ganymede.

\citet{2014SoSyR..48...62M} also studied Jupiter's accretion rates and their compatibility with the Galilean moon system. They found that values between $10^{-9}\,M_{\rm Jup}\,{\rm yr}^{-1}$ and $10^{-6}\,M_{\rm Jup}\,{\rm yr}^{-1}$ (about $10\,M_{\rm Gan}\,\mathrm{Myr}^{-1}$ to $10^{4}\,M_{\rm Gan}\,\mathrm{Myr}^{-1}$) satisfy these constraints. Obviously, our results are at the lower end of this range, while accretion rates up to three orders of magnitude higher should be reasonable according to \citet{2014SoSyR..48...62M}. Yet, these authors also claimed that planetary illumination be negligible in the final states of accretion, which is not supported by our findings (see Fig.~\ref{fig:T_shots_1}). We ascribe this discrepancy to the fact that they estimated the planetary luminosity ($L_{\rm p}$) analytically, while in our model $L_{\rm p}$, $R_{\rm p}$, $M_{\rm p}$, and $\dot{M}$ are coupled in a sophisticated planet evolution model \citep{2013A&A...558A.113M}. In the \citet{2014SoSyR..48...62M} simulations, $10^{-7}\,L_\odot~\lesssim~L_{\rm p}~\lesssim~10^{-4}\,L_\odot$ (depending on assumed values for $R_{\rm p}$, $M_{\rm p}$, and $\dot{M}$), whereas in our case $L_{\rm p}$ remains close to $10^{-4}\,L_\odot$ once a circumjovian H$_2$O ice line forms, while $\dot{M}$ and $R_{\rm p}$ evolve rapidly due to the planet's gap opening in the circumsolar disk \citep[see Fig.~1 in][]{2014arXiv1410.5802H}.

We also calculate the type I migration time scales ($\tau_{\rm I}$) of potential moons that form in our circumplanetary disks. Using Eq.~(1) from \citet{2006Natur.441..834C} and assuming a Ganymede-mass moon, we find that $0.1\,{\rm Myr}~\lesssim~\tau_{\rm I}~\lesssim~100\,{\rm Myr}$ in the final stages of accretion, with the shortest time scales referring to close-in moons and early stages of accretion when the gas surface density is still high. Hence, since the remaining disk lifetime ($\approx10^6$\,yr) is comparable or smaller than the type I migration time scale, migration traps might be needed to stop the moons from falling into the planet. As the gas surface density is decreasing by about an order of magnitude per $10^5$\,yr \citep[see Fig.~1(c) in][]{2014arXiv1410.5802H} and since $\tau_{\rm I}$ is inversely proportional to $\Sigma_{\rm g}$, type I migration slows down substantially even if the protosatellites still grow. On the other hand, if a moon grows large enough to open up a gap in the circumplanetary accretion disk, then type II (inwards) migration might kick in, reinforcing the need for moon migration traps. All these issues call for a detailed study of moon migration under the effects of ice lines and other traps.

Another issue that could cause moon loss is given by their possible tidal migration. If a planet rotates very slowly, its corotation radius will be very wide and any moons would be forced to tidally migrate inwards (ignoring mean motion resonances for the time being). But if the planet rotates quickly and is not subject to tides raised by the star, such as Jupiter, then moons usually migrate outwards due to the planetary tides and at some point they might become gravitationally unbound. \citet[][see their Fig.~2]{2002ApJ...575.1087B} showed that Mars-mass moons around giant planets do not fall into the planet and also remain bound for at least 4.6\,Gyr if the planet is at least 0.17\,AU away from a Sun-like star. Of course, details depend on the exact initial orbit of the moon and on the tidal parameterization of the system, but as we consider planets at several AU from the star, we conclude that loss of moons due to tidal inward or outward migration is not an issue. Yet, it might have an effect on the orbital distances where we can expect those giant moons to be found, so additional tidal studies will be helpful.

\section{Discussion}

If the abundant population of super-Jovian planets at about 1\,AU and closer to Sun-like stars formed in situ, then our results suggest that these planets could not form massive, super-Ganymede-style moons in their accretion disks. These disks would have been too small to feature H$_2$O ice lines and therefore the growth of icy satellites. The accretion disks around these planets might still have formed massive, rocky moons, similar in composition to Io or Europa, which would likely be in close orbits \citep[thereby raising the issue of tidal evolution,][]{2002ApJ...575.1087B,2009ApJ...704.1341C,2011ApJ...736L..14P,2013AsBio..13...18H,2014AsBio..14..798H}, since circumplanetary accretion disks at 1\,AU are relatively small. If these large, close-orbit rocky moons exist, they also might be subject to substantial tidal heating. Alternatively, super-Jovians might have captured moons, e.g. via tidal disruption of binary systems during close encounters \citep{2006Natur.441..192A,2013AsBio..13..315W}, so there might exist independent formation channels for giant, possibly water-rich moons at 1\,AU.

If, however, these super-Jovian planets formed beyond 3 to 4.5\,AU and then migrated to their current locations, then they could be commonly orbited by Mars-mass moons with up to 50\,\% of water, similar in composition to Ganymede and Callisto. Hence, the future detection or non-detection of such moons will help to constrain rather strongly the migration history of their host planets. What is more, Mars-mass ocean moons at about 1\,AU from Sun-like stars might be abundant extrasolar habitats \citep{1997Natur.385..234W,2013AsBio..13...18H,2014AsBio..14..798H}, 

Our results raise interesting questions about the formation of giant planets with satellite systems in the solar system and beyond. The field of moon formation around super-Jovian exoplanets is a new research area, and so many basic questions still need to be answered.

\textbf{(1) The ``Grand Tack''.} In the ``Grand Tack'' scenario \citep{2011Natur.475..206W}, the fully accreted Jupiter migrated as close as 1.5\,AU to the Sun during the first few million years of the solar system, then got caught in a mean-motion orbital resonance with Saturn and then moved outward to about 5\,AU. Our results suggest that the icy moons Ganymede and Callisto can hardly have formed during the several $10^5$\,yr Jupiter spent inside 4.5\,AU to the Sun. If they formed before Jupiter's tack, could their motion through the inner solar system be recorded in these moons today? Alternatively, if Callisto required $10^5$ - $10^6$\,yr to form \citep{2002AJ....124.3404C,2003Icar..163..198M} and assuming that Jupiter's accretion disk was intact until after the tack, one might suppose that Ganymede and Callisto might have formed thereafter. But then how did Jupiter's accretion disk re-acquire the large amounts of H$_2$O that would then be incorporated into Callisto after all water had been sublimated during the tack? We will address these issues in a companion paper \citep{HMP2015}.

\textbf{(2) Migrating planets.} Future work will need to include the actual migration of the host planets, which we neglected in this paper. \citet{2010ApJ...719L.145N} studied the orbital stability of hypothetical moons about migrating giant planets, but the formation of these moons was not considered. Yet, the timing of the accretion evolution, the gap opening, the movement within the circumstellar disk (and thereby the varying effect of stellar heating), and the shutdown of moon formation will be crucial to fully assess the possibility of large exomoons at about 1\,AU. These simulations should be feasible within the framework of our model, but the precomputed planet evolution tracks would need to consider planet migration. Ultimately, magneto-hydrodynamical simulations of the circumplanetary accretion disks around migrating super-Jovian planets might draw a full picture.

\textbf{(3) Directly imaged planets.} Upcoming ground-based extremely large telescopes such as the \textit{E-ELT} and the \textit{Thirty Meter Telescope}, as well as the \textit{James Webb Space Telescope} have the potential to discover large moons transiting directly imaged planets in the infrared \citep{2013ApJ...769...98P,2014ApJ...796L...1H}. Exomoon hunters aiming at these young giant planets beyond typically 10\,AU from the star will need to know how moon formation takes place in these possibly very extended circumplanetary accretion disks, under negligible stellar heating, and in the low-density regions of the circumstellar accretion disk.

\textbf{(4) Ice line traps.} If circumplanetary H$_2$O ice lines can act as moon migration traps and if solids make up a substantial part of the final masses accreted by the planet, then the accretion rates onto giant planets computed under the neglect of moons might be incorrect in the final stages of accretion. The potential of the H$_2$O ice line to act as a moon migration trap is new \citep{2014arXiv1410.5802H} and needs to be tested. It will therefore be necessary to compute the torques acting on the accreting moons within the circumplanetary accretion disk as well as the possible gap opening in the circumplanetary disk by large moons, which might trigger type II migration \citep{2002AJ....124.3404C,2006Natur.441..834C}.

\section{Conclusions}

Planetary illumination is the dominant energy source in the late-stage accretion disks around Jupiter-mass planets at 5.2\,AU from their Sun-like host stars (Fig.~\ref{fig:T_shots_1}), while viscous heating can be comparable in the final stages of accretion around the most massive planets (Fig.~\ref{fig:T_shots_12}).

At the time of moon formation shutdown, the H$_2$O ice line in accretion disks around super-Jovian planets at 5.2\,AU from Sun-like host stars is between roughly 15 and $30\,R_{\rm Jup}$. This distance range is almost independent of the final planetary mass and weakly dependent on the disk's absorption properties (Fig.~\ref{fig:stats}). With more massive planets having more extended accretion disks, this means that more massive planets have larger fractions (up to $70$\,\%) of their disks beyond the circumplanetary H$_2$O ice line (compared to about 25\,\% around Jupiter, see Fig.~\ref{fig:statsfrac}).

Jupiter-mass planets forming closer than about 4.5\,AU to a Sun-like star do not have a circumplanetary H$_2$O ice line (Fig.~\ref{fig:distance_1Mjup}), depending on the opacity details of the circumstellar disk. A detailed application of this aspect to the formation of the Galilean satellites might help constraining the initial conditions of the Grand Tack paradigm. Due to their larger disks, the most massive super-Jovian planets can host an H$_2$O ice line as close as about 3\,AU to Sun-like stars (Fig.~\ref{fig:distance_12Mjup}). With the circumstellar H$_2$O ice line at about 2.7\,AU in our model of an optically thin circumstellar disk \citep{1981PThPS..70...35H}, the relatively small accretion disks around Jupiter-mass planets at several AU from Sun-like stars thus substantially constrain the formation of icy moons. The extended disks around super-Jovian planets, on the other hand, might still form icy moons even if the planet is close to the circumstellar H$_2$O ice line.

We find an approximation for the total mass available for moon formation ($M_{\rm T}$), which is a function of both $M_{\rm p}$ and $r_\star$ (see Eq.~\ref{eq:M_T}). The linear dependence of $M_{\rm T}~\propto~\,M_{\rm p}$ has been known before, but the dependence on stellar distance ($M_{\rm T}~\propto~\,r_\star^{1.16}$, for $r_\star~\geq~5.2$\,AU) is new. It is based on our finding that an accretion rate $M_{\rm shut}~\approx~10\,M_{\rm Gan}/10^{6}$\,yr (Fig.~\ref{fig:MsInst}) yields the best results for the position of the circumjovian H$_2$O ice line at the shutdown of the formation of the Galilean moons (Fig.~\ref{fig:T_shots_1}b); and it assumes that this shutdown accretion rate is similar around all super-Jovian planets.

Our results suggest that the observed large population of super-Jovian planets at about 1\,AU to Sun-like stars should not be orbited by water-rich moons if the planets formed in-situ. However, in the more plausible case that these planets migrated to their current orbits from beyond about 3 to 4.5\,AU, they should be orbited by large, Mars-sized moons with astrobiological potential. As a result, future detections or non-detections of exomoons around giant planets can help to distinguish between the two scenarios because they are tracers of their host planets' migration histories.

\begin{acknowledgements}
We thank S{\'e}bastien Charnoz for his referee report which helped us to clarify several passages of this manuscript. We thank Christoph Mordasini for providing us with the precomputed planetary evolution tracks. Ren\'e Heller is supported by the Origins Institute at McMaster University and by the Canadian Astrobiology Training Program, a Collaborative Research and Training Experience Program funded by the Natural Sciences and Engineering Research Council of Canada (NSERC). Ralph E. Pudritz is supported by a Discovery grant from NSERC. This work made use of NASA's ADS Bibliographic Services and of The Extrasolar Planet Encyclopaedia (\url{www.exoplanet.eu}). Computations have been performed with {\tt ipython 0.13.2} on {\tt python 2.7.2} \citep{PER-GRA:2007}, and all figures were prepared with {\tt gnuplot 4.6} (\href{http://www.gnuplot.info}{www.gnuplot.info}).
\end{acknowledgements}


\bibliographystyle{aa} 
\bibliography{ms}




\end{document}